\documentclass[twocolumn]{aastex631}
\usepackage[mathlines]{lineno}
\usepackage{apjfonts}
\usepackage{natbib}
\usepackage{xspace}
\usepackage{booktabs}
\usepackage{amsmath}
\usepackage{color}
\usepackage[utf8]{inputenc}
\usepackage{rotating}
\usepackage{url}
\usepackage[frozencache=true, cachedir=.]{minted}

\usepackage[english]{babel}

\newcommand{\kepler}{\emph{Kepler}\xspace}
\newcommand{\ktwo}{\emph{K2}\xspace}
\newcommand{\tess}{\emph{TESS}\xspace}
\newcommand{\gaia}{\emph{Gaia}\xspace}
\newcommand{\lk}{\texttt{lightkurve}\xspace}

\newcommand{\koi}{\textit{KOI-608}\xspace}
\newcommand{\nsources}{576\xspace}

\newcommand{\acronym}{Linearized Field Deblending\xspace}
\newcommand{\acnm}{LFD Photometry\xspace}

\newcommand{\gsourcea}{\textit{Gaia EDR3 2073618323728332544}\xspace}
\newcommand{\gsourceb}{\textit{Gaia EDR3 2073618220649116160}\xspace}

\shortauthors{Hedges et al.}

\makeatletter
\newcommand\footnoteref[1]{\protected@xdef\@thefnmark{\ref{#1}}\@footnotemark}
\makeatother

\begin{document}
\title{Linearized Field Deblending: PSF Photometry for Impatient Astronomers}

\email{christina.l.hedges@nasa.gov}

\author{Christina Hedges}
\affil{Bay Area Environmental Research Institute, P.O. Box 25, Moffett Field, CA 94035, USA}
\affil{NASA Ames Research Center, Moffett Field, CA}

\author{Rodrigo Luger}
\affil{Center for Computational Astrophysics, Flatiron Institute, New York, NY}

\author{Jorge Martinez Palomera}
\affil{Bay Area Environmental Research Institute, P.O. Box 25, Moffett Field, CA 94035, USA}
\affil{NASA Ames Research Center, Moffett Field, CA}

\author{Jessie Dotson}
\affil{NASA Ames Research Center, Moffett Field, CA}

\author{Geert Barentsen}
\affil{Bay Area Environmental Research Institute, P.O. Box 25, Moffett Field, CA 94035, USA}
\affil{NASA Ames Research Center, Moffett Field, CA}


\begin{abstract}

NASA's \kepler, \ktwo and \tess missions employ Simple Aperture Photometry (SAP) to derive time-series photometry, where an aperture is estimated for each star, and pixels containing each star are summed to create a single light curve. This method is simple, but in crowded fields the derived time-series can be highly contaminated. The alternate method of fitting a Point Spread Function (PSF) to the data is able to account for crowding, but is computationally expensive. In this paper, we present a new approach to extracting photometry from these time-series missions, which fits the PSF directly, but makes simplifying assumptions in order to greatly reduce the computation expense. Our method fixes the scene of the field in each image, estimates the PSF shape of the instrument with a linear model, and allows only source flux and position to vary. We demonstrate that our method is able to separate the photometry from blended targets in the \kepler dataset that are separated by less than a pixel. Our method is fast to compute, and fully accounts for uncertainties from degeneracies due to crowded fields. We name the method described in this work \acronym (\acnm). We demonstrate our method on the false positive \kepler target \koi. We are able to separate the photometry of the two sources in the data, and demonstrate the contaminating transiting signal is consistent with a small, sub-stellar companion with a radius of $2.67R_{jup}$ ($0.27R_{sol}$). Our method is equally applicable to extracting photometry from NASA's \tess mission.

\end{abstract}

\section{Introduction}

NASA's \kepler mission provided photometry of over 400,000 stars \citep{kepler}, revolutionizing the field of study of transiting exoplanets. \kepler relied on the \kepler Pipeline to extract the photometry of these sources \citep{pipeline}, which provided high quality data products including 1) Target Pixel Files (TPFs) 2) Simple Aperture Photometry (SAP) light curves 3) Pre Data-search Conditioning Simple Aperture Photometry (PDCSAP) light curves \citep{pdcsap}. TPFs consist of stacks of exposures containing pixel cut-outs around a single star, usually covering $\approx5^2$ pixels on the detector. SAP light curves consist of time-series photometry using pipeline generated apertures.  PDCSAP light curves are the SAP light curves, corrected for instrument systematics using Contrending Basis Vectors (CBVs) \citep[see][]{cbvs}. NASA's \tess mission recently completed its prime mission, observing over 50M stars \citep{tess, tesstom} and uses similar methods to the \kepler pipeline to produce light curves.

In this paper, we present a new approach to extract photometry from datasets similar to those from the \kepler mission, which we name \acronym (\acnm). Rather than using SAP methods, we develop a method of modeling the astrophysical scene, where we create a simple, fast to evaluate model for the Pixel Response Function (PRF) of each source on the detector, and rely on the \gaia catalog to correctly place each source in the scene, fitting only for the flux in each source. We then simultaneously fit all sources in the image, which enables us to accurately account for crowding and contamination in the time-series of each source. Our method employs a linear model, and is able to accurately "deblend" crowded fields, fully accounting for uncertainties due to the crowding in the scene, and so we name our method \acronym. \acnm is fast, (in this work we present a demonstration fitting \nsources sources simultaneously in under 4 minutes on a personal laptop), requires few free parameters, and provides reasonable photometric errors, fully accounting for uncertainties due to crowding. As such, we suggest \acnm is ideal for the impatient time domain astronomer.

In Section~\ref{sec:photometry} of this paper we discuss the different approaches to photometry in literature, and the benefits of scene modeling. In Section~\ref{sec:improvements} we discuss key decisions we have made to make our method more tractable. In Section~\ref{sec:implementation} we present our model, in the context of a demonstration target: \koi. In Section~\ref{sec:discussion} we discuss the limitations of our model, and the potential applications.


\section{Background: Photometry}
\label{sec:photometry}

Performing high-precision photometry for an astronomical source from instruments like Kepler requires identifying which pixels contain signal for the source, and combining them to maximize the signal to noise and precision of the photometry. The most common general purpose methods are aperture photometry and PSF photometry. Difference imaging is an alternative method used when the primary goal is to identify variable sources. The \kepler mission itself employed aperture photometry, which is also favored by ground-based observatories. In this section we briefly discuss common photometry methods.

\subsection{Aperture Photometry}

Aperture photometry is a process of selecting pixels to be summed in a stack of images, in order to create time series photometry. Simple Aperture Photometry (SAP) sums all the pixels assigned to each source to create a single photometric time series. This method benefits from being simple, and fast to apply to targets. However, in cases of severe crowding, it can be difficult to assign apertures, and a trade must be made between completeness of the source flux and contamination from neighbors.

The \kepler pipeline utilized simple aperture photometry with excellent results\citep{Kepler-PA}.  Two methods were used to identify candidate pixels for each source, one based on a model of the image and one based onthe raw pixel data.  The final aperture was chosen to maximize SNR and minimize the combined photometric precision (CDPP) for each source \citep{Kepler-OA}. Crowding and contamination metrics were also provided.

Other methods use a modified version of this approach by assigning weights to each pixel in the aperture. For example, the  \texttt{photutils} package enables users to build apertures from "partial" pixels by assigning pixel level weights, enabling users to build, for example, circular apertures. Another example of a weighted aperture method is the method introduced by \citep{1998MNRAS.296..339N}, where the pixel weights are computed from a PSF model profile fitted to stars in the vicinity and scaled by the variance of each pixel. This method offers robust error estimations and, assuming that the PSF profile does not vary significantly across sub-regions of the image, computing times can be tuned by fitting a lower number of PSF models in an image \citep{2016ApJ...832..155F}.

Several tools can be used to create aperture photometry, for example \texttt{photutils} provides general Python libraries for aperture photomtery. The \texttt{lightkurve} package \citep{lightkurve} can be used to create simple aperture photometry from \kepler, \ktwo, and \tess. The \texttt{eleanor} package \citep{eleanor} performs both simple aperture photometry and weighted aperture photometry on sources observed in \tess full frame images using a set of predefined simple and weighted apertures.  Whether simple or weighted, aperture photometry usually provides a fast way to create time-series photometry once pixels have been assigned to apertures. Usually some component of scene modeling or point finding is required to assign pixels to apertures.


\subsection{PSF Photometry}

\begin{figure}
    \centering
    \includegraphics[width=0.5\textwidth]{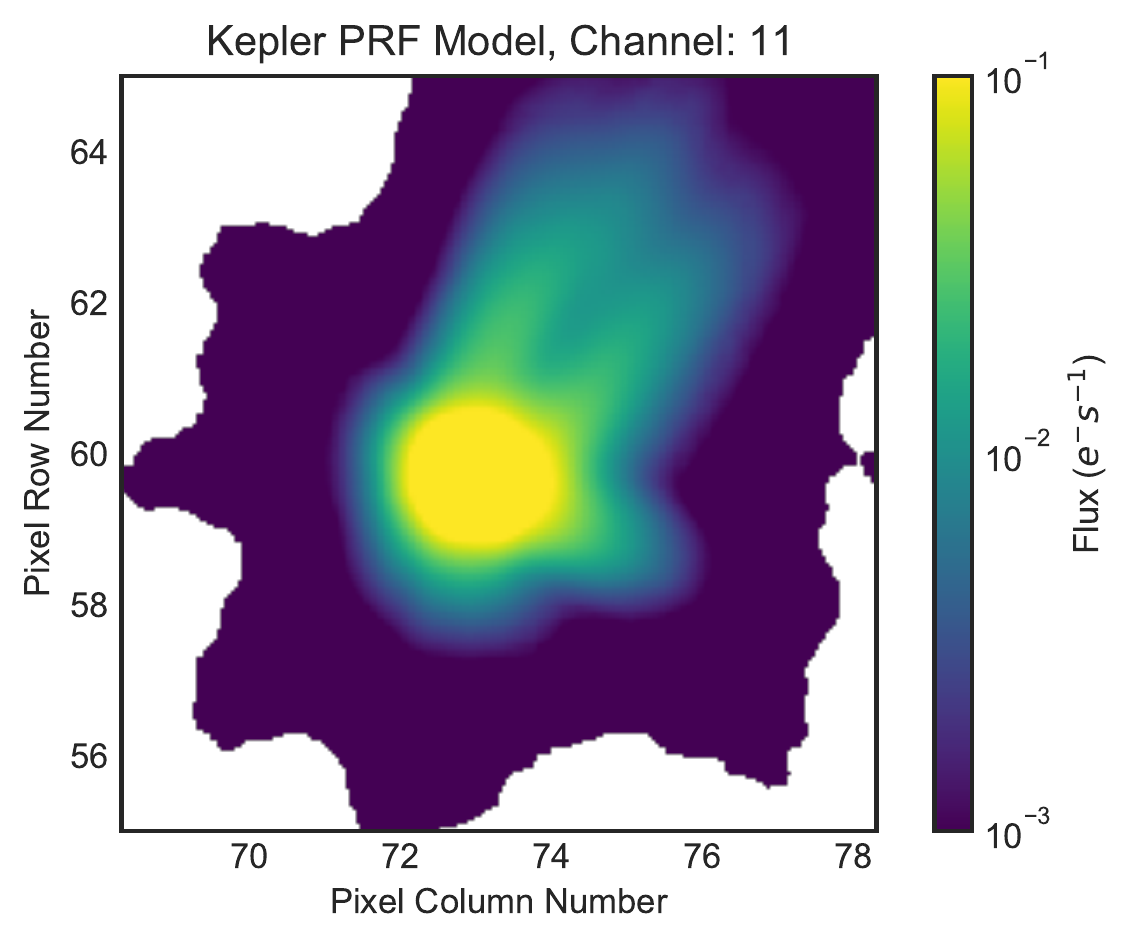}
    \caption{Example Pixel Response Function for \kepler on Channel 11, developed by \cite{bryson} and evaluated using \lk \citep{lightkurve}. The PRF is an estimate of the Point Spread Function of the telescope, after it has been convolved with the instrument systematics and integrated on the pixel grid. White portions of this figure indicate where the original model is either undefined, or has negative values.}
    \label{fig:prf}
\end{figure}

In PSF photometry, an estimate of the PSF shape is used, and both the flux and the position of the source are fit, either simultaneously or separately. \cite{anderson2000} and \cite{anderson2006} demonstrate how to build a model for the ``effective" PSF (ePSF) of an instrument, and fit to retrieve the flux and position of sources for HST instruments. This ePSF is similar to the \kepler Pixel Response Function (PRF) presented in \cite{bryson}, which they define as the optical Point Spread Function of the \kepler instrument, having been convolved with the instrument systematics, pixel sensitivities, and any intra-cadence motion, and having been recorded on the detector Figure~\ref{fig:prf} shows an example of the \kepler PRF, created using dithered data. Whether using some estimate of the instrument PSF, or a data-driven model of the ePSF/PRF, this method usually fits for both position and flux.

Several works have used PSF or PRF photometry on \kepler data. For example \cite{nardiello} derive PSF models, fit the flux and positions of stars neighbouring their target star, remove these neighbors, and calculate the target photometry both using PSF and aperture photometry. Similarly, \cite{libralato} estimate the Pixel Response Function of the \kepler instrument and perform the same fit. Open source tools such as \texttt{photutils}\footnote{See \url{https://photutils.readthedocs.io/en/stable/psf.html}} \citep{photutils} have enabled users to perform PSF photometry, and reimplement algorithms from DAOPHOT \citep{daophot} in Python. Because PSF photometry usually fits both position and flux, and fits each star independently, it can sometimes be difficult in practice to constrain a model. Owing to these many parameters, PSF photometry is also much slower to compute than SAP.

Some works implement a simplified, less flexible verison of PSF photometry. For example, \cite{eleanor} implements a simplified PSF photometry option on \tess data (which is similar in format to \kepler data). The locations of the stars are assumed to be known apriori.  The PSF is assumed to be a 2-dimensional gaussian.  The two widths and orientation of the PSF is fit from an ensemble of sources.  The height of the PSF is allowed to vary for each source.  While computationally less intensive than other PSF implementations, this approach is not well suited to crowded regions due to the assumption of a gaussian PSF, and a static scene.

\subsection{Difference imaging}


Difference imaging, or image subtraction, consists of aligning the images in a stack of observations to a reference frame, accounting for differences in seeing or PSF shape through kernel convolution, and subtracting images to find the difference. Difference imaging was first discussed in this way in \cite{diff1}, and kernels were later improved in \cite{diff2} and \cite{diff3}. Difference imaging has been applied to the \kepler/\ktwo mission to extract light curves from extremely crowded fields such as the galactic center in \cite{wang2017} (who developed a Causal Pixel Modeling approach to difference imaging) and \cite{k2c9}, and to open clusters in \citep{k2c0}. This approach can be useful when looking for transient events, which appear as high signal to noise events in difference imaging where quiescent phenomena do not. However, difference imaging usually provides only a difference in flux, and therefore requires external information to retrieve the flux of a source in absolute terms. Additionally, difference imaging can be highly sensitive to the precision of the image registration.

\subsection{\acnm}
In this work we approach the problem by attempting to recreate the entire image of all stars at the pixel level, simultaneously. In the PSF/PRF photometry approach, catalogs are often used to build the scene or build a starting point, but in most cases this input position is allowed to vary, and stars are fit either individually or in small groups. In \acnm we instead 1) fix the locations of each source 2) predict the flux in each source.

In the method we discuss in this work, a pre-existing catalog is used that is more accurate than the pixel scale of the detector. We use \gaia EDR3, which provides extremely precise and complete locations for targets in \kepler's sensitivity range. We build a PRF model for all sources, evaluate this model at the location of each source given by \gaia, and fit for the flux of each source simultaneously in an individual image. This processes is repeated on a stack of images in order to obtain the time series. To account for common motion of the ``scene" (e.g. due to velocity aberration) we may allow the model to drift over time (see Section~\ref{sec:motion}). We use \gaia G band photometry as an estimate for the expected flux observed with \kepler, which provides the best match to the \kepler bandpass of any similarly complete survey. We assume that Gaia is complete to a \kepler magnitude of 18, and that any stars unresolved by \gaia due to crowding are indistinguishable from single stars in \kepler.

\acnm is less flexible than PSF/PRF modeling, as we are not able to vary the positions of the sources independently. This can be seen as a strength of the method, as this prevents degeneracies if, for example in the case of \kepler, a source is moving out of a TPF. In such a case, it can be difficult to distinguish between a source moving, or a source changing flux. In \acnm, no source is allowed to move independently from the other sources, and so we are able to break this degeneracy. Since all sources are fit simultaneously, we can retrieve the correct errors, marginalized over the uncertainty due to crowding. This approach is also faster than PSF/PRF photometry for a large number of sources, since there are fewer free parameters (i.e. no position parameters). We propose modeling the whole astronomical scene using our method is a highly useful tool in the era of \kepler and \tess.

In this work, we will demonstrate how to fit the PRF shape of a given stack of well aligned images to use in \acnm (see Section~\ref{sec:implementation}). In our specific case of \kepler data, it would also be possible to utilize the measured \kepler PRF from \cite{bryson}, which was built from all sources on a single \kepler channel. In our method we are essentially re-deriving a slightly noisier estimate of the PRF, by creating a model of the PRF using fewer sources, and allowing no flexibility in the shape as a function of position on the detector. We choose to do this so that we are able to 1) obtain the best fitting, local PRF 2) obtain the polar coordinate model for the PRF, which is fast to compute and evaluate 3) demonstrate the general approach here which is equally applicable to \kepler, \ktwo, \tess, and beyond. In theory, it is possible to fit existing models in the ways described in this work (e.g. from \cite{bryson}, or from optical simulations) and use those to fit source weights, rather than building a data driven model.

\subsection{Detrending}

Many light curves, both from the \kepler and \tess missions and from ground-based observatories, use postprocessing to correct instrumental effects. For example, the CBV method corrects motion in \kepler data due to velocity aberration, and corrects for focus change due to spacecraft heating \citep{cbvs}. Methods such as the Self Flat Fielding (SFF) technique \citep{vanderburg} and Pixel Level Decorrelation \citep{luger} have been used to post-process SAP light curves from the \ktwo mission (the second phase of the \kepler mission) to remove the high frequency roll motion of the telescope. \cite{wang2017} used Causal Pixel Modeling in combination with a difference imaging approach to detrend \ktwo the spacecraft motion in the Campaign 9 Microlensing campaign. \cite{Aigrain2015} employed a Gaussian Process to model and detrend the \ktwo roll motion in light curves derived using a circular aperture. These varied post processing approaches can improve the photometric precision of light curves generated using apertures, PSFs, or difference imaging. In this work, we demonstrate how to extract photometry, but attempt no further postprocessing. However, the CBV detrending and SFF techniques to post process SAP light curves are just as applicable to light curves derived using \acnm, and will similarly improve photometric precision.

\section{Simplifying Assumptions Employed in LFD}
\label{sec:improvements}


In order to compute our model of the image quickly, we require a simple, spatial model of the PRF that is efficient to fit and evaluate. To do this, we require simplifying assumptions. In \acnm, we undertake two key simplifying assumptions, described below.

In our method, we design a simple linear model of low order, piece-wise polynomials. PSFs (and thus PRFs) tend to have high dynamic range (tall peaks, and wide, shallow wings) which can make fitting a linear model difficult. PSFs are also narrow, usually covering just a few pixels. Other tools, such as \texttt{photutils}, create a fine grid of subpixels, and then coadd many PSFs from the data on this fine (oversampled) grid. This finer grid can then be fit with some model, or simply interpolated. However, fitting a linear model to this gridded data in order to evaluate it anywhere on the detector can still difficult and expensive. Instead, we take the following approaches to simplify the problem.

\subsection{Fitting in polar coordinates}

\begin{figure}
    \centering
    \includegraphics[width=0.5\textwidth]{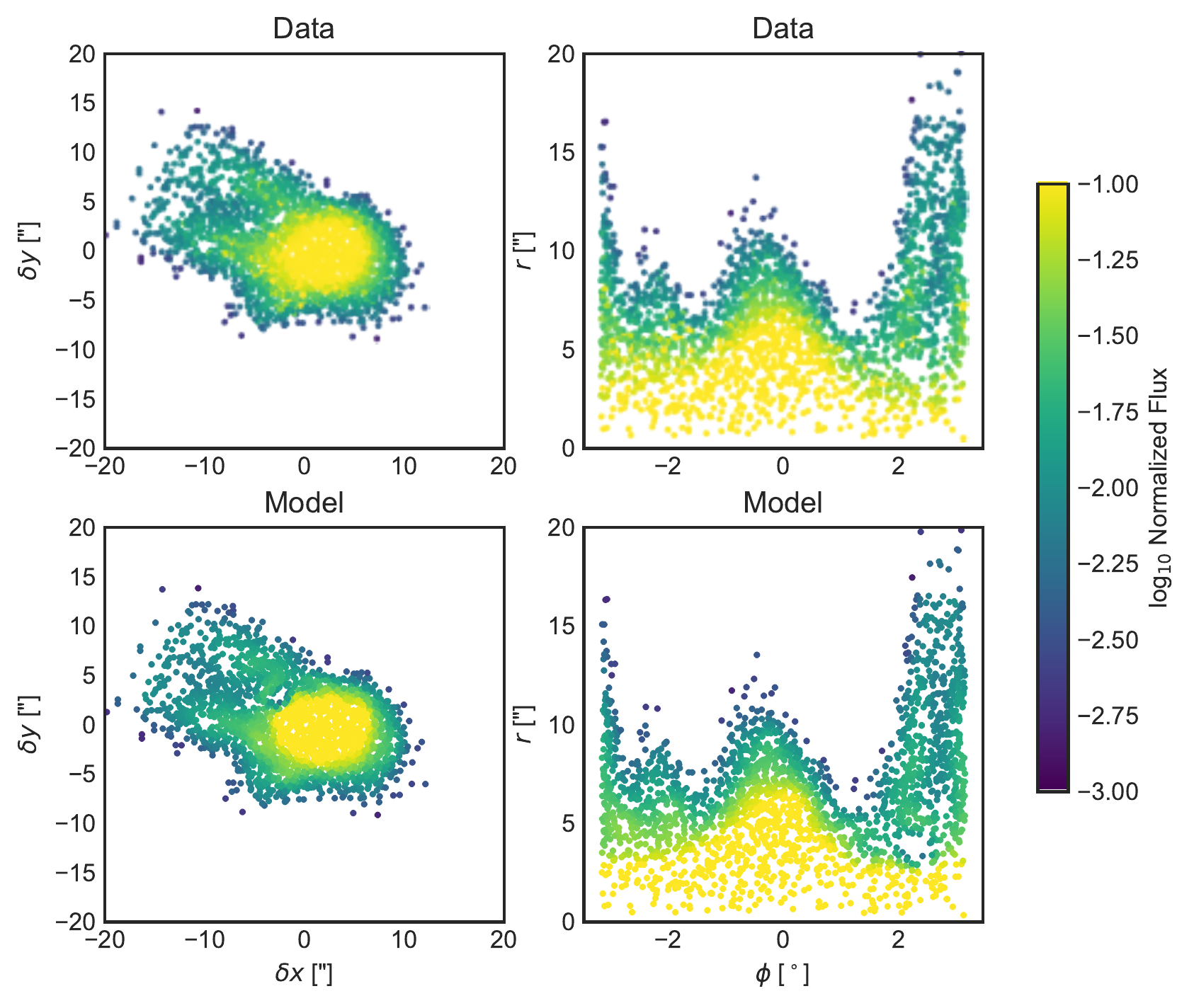}
    \caption{PRF based on data from NASA's \kepler mission from Quarter 10 and Channel 11, and our model. \nsources distinct sources are each normalized by their total flux and then plotted in both Cartesian space (first panel) and polar coordinate space (second panel). Each point represents a pixel value from a single cadence of \kepler data, the color of each point represents the normalized flux value. In Cartesian space the data has steep gradients, but in polar coordinate space the data varies smoothly, making it easier to model with a simple set of basis-splines. The third panel shows our model in polar coordinates, and the final panel shows the polar coordinate model, recast back into Cartesian coordinates, which has good agreement with the data and preserves the PSF wings. $\delta x$, $\delta y$, and $r$ are in units of pixels, and $\phi$ is in units of radians. Note that in Cartesian coordinates, the PRF measured here is highly similar to the one measured by \cite{bryson} shown in Figure~\ref{fig:prf}.
    }
    \label{fig:psfdemo1}
\end{figure}

PSFs (and so PRFs) usually have some degree of natural radial symmetry, even in cases where they are elongated in one direction. Rather than fitting in Cartesian coordinates, where the PSF changes rapidly on small spatial scales, fitting in polar coordinates allows us to benefit from this radial symmetry, and fit a smooth model. In Cartesian space we would fit the flux of a PSF as a function of two parameters,

\begin{equation}
    \delta x = x - x_0
\end{equation}
\begin{equation}
    \delta y = y - y_0
\end{equation}
where $x$ and $y$ denote the pixel position in each dimension, and $x_0$ and $y_0$ denote the center of a given source. $\delta x$ and $\delta y$ indicate the separation from the source in the $x$ and $y$ dimensions. Instead, in this work we fit the flux as a function of

\begin{equation}
    \label{eq:pol1}
    r = (\delta x^2 + \delta y^2)^{\frac{1}{2}}
\end{equation}
\begin{equation}
    \label{eq:pol2}
    \phi = \textrm{arctan2}\Big(\delta y, \delta x\Big)
\end{equation}
where $r$ is the radial distance from the PSF center and $\phi$ is the azimuth angle around the PSF. The top panels of Figure~\ref{fig:psfdemo1} show an example of data from the \kepler mission in each of these coordinate systems, demonstrating that the flux data from the telescope varies more smoothly in $\phi$ and $r$ than in $\delta x$ and $\delta y$.

\subsubsection{Fit in log of flux}

PSFs have a large dynamic range, and often have wings that are orders of magnitude fainter than their peaks. In the case of \kepler, channels at the edge of the focal plane have PSFs with highly extended wings \citep[see the Kepler Instrument Handbook, hereafter referred to as KIH][]{kih}. As shown in Figures~\ref{fig:prf} and ~\ref{fig:psfdemo1}, the wings of the PRF are up to 2 orders of magnitude fainter than the center of the PRF. The flux varies steeply as a function of radial distance from the center of the PSF, following a power-law (approximately $1/r^2$). As a simplifying assumption, we fit in the log of the flux, where-by we can fit this power law with a simple polynomial, meaning that we can create a simple linear model. We additionally benefit from log space enforcing that the model must always be positive.


\subsection{Basis splines}

In this work, we use basis splines to model the PRF. Basis splines are nonparameteric, and model data as a simple, piece-wise polynomials, which makes our PRF model linear and therefore fast to optimize and evaluate. A basis spline is defined by a number of ``knots", which are points where the piece-wise polynomials meet. A basis spline is defined such that, if we specify a spline with degree of 3, the first two derivatives of the function at these knots are continuous, i.e. at these knots, the piece-wise polynomials must vary smoothly. We can create a design matrix of basis spline components that is prescribed by a number of knots in the $\phi$ and $r$ dimensions, which controls the level of detail we are able to capture in our model. Increasing the number of knots increases the model complexity, and increases the computation time. We describe a basis spline in detail in Appendix~\ref{sec:bsplines}.


As we will use polar coordinates when fitting the PRF model, our basis spline must ``wrap" in the $\phi$ dimension, (i.e., the value at 0$^\circ$ must be identical to the value at 360$^\circ$). We include a simple implementation of how to create such a basis spline using Python in the Appendix of this work.

To create a basis spline model we create a design matrix, made up of vectors containing each part of the piece-wise polynomial for a given variable. When this matrix is dotted with a vector of weights, we retrieve a smoothly varying model. See Appendix ~\ref{sec:linearappendix} for the general example of how to fit to find the best fitting weights for the design matrix using linear least squares. In the sections below, we use the equations in Appendix ~\ref{sec:linearappendix} to fit for weights using several different design matrices.






\section{Implementation}
\label{sec:implementation}
Our approach to model a stack of astronomical images from \kepler follows these stages:
\begin{enumerate}
    \item Obtain the source positions and average \textit{G} band brightness in the scene using \gaia.
    \item Build an estimate of the edge of the PRF, in order to identify pixels that have a contribution from a single source ("uncontaminated" pixels).
    \item Build a linear model for the PRF shape using uncontaminated pixels, which can be evaluated quickly at a given separation from a source center. Iteratively fit and mask out data to build an accurate model of PRF shape.
    \item Extract pixel time series of uncontaminated pixels, mean normalize, and model the change in flux due to common motion in the scene as a function of time.
    \item Model the image stack as a combination of the PRF shape model and the common motion model, where the weight for each source as a function of time gives the flux time-series of the source.
\end{enumerate}
These steps are described in detail below. In Section~\ref{sec:flowchart} we conclude with a diagram of the steps involved in our method.

We make the following implicit assumptions:

\begin{itemize}
    \item Our catalog (in this case \gaia \emph{EDR3}) is accurate, and we can ignore any errors in the location of each source in the scene. Extending this, we assume the World Coordinate System solution for the image stack is accurate.
    \item There is no distortion of the image due to e.g. the optics of the telescope, such that PRF shapes do not vary greatly over the spatial scale of the images.
    \item The motion in the scene is small, less than the width of the PSF, and smooth over time.
    \item The motion in the scene is the same across all sources.
    \item The intra- and inter-pixel sensitivity changes are small, and average out across many sources.
\end{itemize}

We will discuss the limitations of these assumptions in the context of \kepler, \ktwo and \tess in Section~\ref{sec:discussion}. A table with descriptions of all the variables used in this section is given in the Appendix in Table~\ref{sec:table}.

In this section we discuss our implementation of the steps above, in the context of \kepler Object of Interest \koi. This object is a false positive planet candidate from the \kepler mission, that shows a significant transiting signal. The \kepler data of \koi contains two sources, separated by just over a pixel on the detector. The first source (\gsourcea) has a \gaia \textit{g} magnitude of 16.855, and hosts a transiting signal. The second source (\gsourceb) has a \gaia \textit{g} magnitude of 14.723, and was the target of the \kepler survey. Owing to the detection of a significant centroid shift, indicating the signal originated from a faint background source, \koi was discounted as a false positive, and not investigated further. However, in this work we will demonstrate that our PSF photometry method is able to separate these two sources, and show that the transiting signal in \koi is still consistent with being a small, sub-stellar companion. While we use \koi as a demonstration, our approach will work for any \kepler target. We summarise the steps in this section in a flow chart in Section~\ref{sec:flowchart}.

\subsection{Data Preparation}

We obtain the \kepler Pipeline processed Target Pixel File (TPF) data from the Mikulski Archive for Space Telescopes (MAST)\footnote{https://archive.stsci.edu/kepler/} for \koi, and downloaded TPFs of 200 sources neighboring sources. We find in this case that 200 TPFs provides a reasonable balance between having enough sources to constrain the model, while being close enough on the detector to have a similar PSF shape. In this proof of concept, we use only a single \kepler quarter. For each TPF we query \gaia \emph{EDR3} \cite{gaiaedr3} to obtain a catalog of all sources that were observed in the TPFs, down to a limiting magnitude of 18 in \gaia \textit{g} magnitude. We will denote that TPFs have $n$ pixels, and $m$ cadences, and that $l$ unique sources are observed. In our example, $n=6978$ pixels, $m=4442$ cadences, and $l=\nsources$ sources. We will label the image stack data $\mathbf{f}$, which is a matrix with shape ($n$, $m$).

In this work we have matrices of quantities (for example, Cartesian separation from each source), which are 2D and have dimensions ($l,  n$) (number of sources by number of pixels). To use these quantities in our framework, we often "unwrap" these 2D matrices into 1D vectors. We build vectors containing the Cartesian separation from each source, $\pmb{\delta x}$ and $\pmb{\delta y}$, which have length $l \times n$ (number of sources multiplied by number of pixels). In this work, we use the notation $\textrm{vec}(...)$ to denote "unwrapping" a matrix, and $\textrm{mat}(...)$ to denote the inverse (e.g. converting a vector of length $l \times n$ to a 2D matrix of shape ($l$, $n$)). Using Equations~\ref{eq:pol1} and ~\ref{eq:pol2} we create matrices $\pmb{\phi}$ and $\mathbf{r}$ with length $l \times n$.

\subsection{Estimating the edge of the PRF, as a function of flux}
\label{sec:edge}
In order to be efficient in future steps, we first estimate a circular aperture for each source. This initial aperture enables us to identify pixels that are likely to contain only flux from a single source, which are pixels we can use to inform our PRF model. A reasonable circular aperture is neither too small (which omits flux at the edge of the PRF) nor to big (which causes us to over estimate crowding). To find a reasonable circular aperture as a function of flux, we build a basic model of the PRF shape. We then use this model to find the radius at which the recorded counts from a source would reach the noise limit of \kepler. We define this as the edge of the PRF. 

To find the edge of the PRF, we fit the time-averaged (mean) flux of all frames, which we will denote as $\mathbf{\bar{f}}$, and is a vector with length $n$. We model $\mathbf{\bar{f}}$ as a simple polynomial in radial separation from each source $\mathbf{r}$ (which is a vector with length $l \times n$), and our a priori estimate of the source flux. In this initial step, we will use the \gaia mean flux (\texttt{phot\_g\_mean\_flux} in the \gaia EDR3 catalog) as our a priori flux estimate of each source. We will denote this estimate as $\mathbf{g}$, which is a vector with length length $l \times n$, where the value of $\mathbf{g}$ is the same for all pixels for a given source).

We build a design matrix for this process using

\begin{equation}
    \mathbf{X}' = \textrm{vec} \left(
    \begin{bmatrix}
    1\\
    \mathbf{g}\\
    \mathbf{g}^2\\
    \end{bmatrix}
    \begin{bmatrix}
    1 & \mathbf{r} & \mathbf{r}^2
    \end{bmatrix}
    \right)
\end{equation}
\begin{equation}
    \mathbf{X}' =
    \begin{bmatrix}
    1 &
    \mathbf{g} &
    \mathbf{g}^2 &
    \mathbf{r} &
    \mathbf{gr} &
    \mathbf{g}^2\mathbf{r} &
    \mathbf{r}^2 &
    \mathbf{gr}^2 &
    \mathbf{g}^2\mathbf{r}^2 &
    \end{bmatrix}
\end{equation}
where $\mathbf{X}'$ is the design matrix for this initial step. $\textrm{vec(x)}$ denotes the vectorisation operation, which unrolls a matrix into a vector.  $\mathbf{X}'$ is a 2D matrix and has shape ($l \times n$, 9). Using the equations given in Appendix~\ref{sec:linearappendix} we then fit to find the best fitting model flux for every pixel, which we will denote as $\mathbf{\hat{f}_0}$ and is a vector of length $l \times n$. The hat symbol in this work denotes our best mean estimate of a quantity.

\begin{equation}
    \label{eq:edge}
    \mathbf{\hat{f}}_0 = \mathbf{X}'\cdot\mathbf{\hat{w}_0}
\end{equation}
where the subscript $_0$ denotes the model flux for the first stage simple, circular aperture. $\mathbf{\hat{w}}_0$ is a vector of the best fit coefficients for the circular aperture model, with length 9. Equation~\ref{eq:edge} models the data as a simple circular PRF, with dependence only on source flux and radius from the source. We then consider the edge of the PRF to be the radius at which this model is greater than some threshold. We find empirically that a value of 50$e^-s^{-1}$ counts balances reasonably sized apertures that include the majority of the wings of the PSF. This allows us to place simple apertures around every source, and identify "uncontaminated pixels" (pixels with contributions from a single source). Using this model we identify 3418 pixels that are uncontaminated in our demonstration example. 


\subsection{Building a linear model of the PSF shape}
\label{sec:linearmodel}

We use basis splines in polar coordinates to build a model for the shape of the PRF of the \kepler telescope. This model is simple, and so can be solved with linear least squares, yet is powerful enough to completely capture the PSF shape. To employ basis splines, we build a design matrix of basis spline components based on the vectors $\pmb{\phi}$ and $\mathbf{r}$, which have shape $l \times n$. We denote these matrices as $\pmb{\Phi}$ and $\mathbf{R}$ respectively. We build a spline design matrix $\pmb{\Phi}$ with degree 3 and 15 knots, evenly spaced between $-\pi$ and $\pi$. This spline is "wrapped", such that the function has no discontinuity as a function of angle (see Appendix ~\ref{sec:wrap}). We build a 3rd degree spline design matrix $\mathbf{R}$ with 12 knots linearly spaced in radius squared, between a radius of 1" and 18" (approximately 1/4 of a pixel and 4.5 pixels). $\pmb{\Phi}$ has $j_\phi$ components, making it a matrix of shape ($l \times n$, $j_\phi$), and $\mathbf{R}$ has $j_r$ components, making it a matrix of shape ($l \times n$, $j_r$). Examples of the structure of $\pmb{\phi}$ and $\mathbf{r}$ are given below. The values of $j_\phi$ and $j_r$ tune the resolution of the shape model, and depend on the number of knots and degree of the b-spline.

\begin{equation}
    \pmb{\Phi} = \begin{bmatrix}
    \pmb{\Phi}_{0, 0} & \pmb{\Phi}_{1, 0} & ... & \pmb{\Phi}_{j_\phi, 0} \\
    \pmb{\Phi}_{0, 1} & \pmb{\Phi}_{1, 1} & ... & \pmb{\Phi}_{j_\phi, 1} \\
    ... & ... & ... & ... \\
    \pmb{\Phi}_{0,\; n \times l} & \pmb{\Phi}_{1,\;n \times l} & ... & \pmb{\Phi}_{j_\phi,\; n \times l} \\
    \end{bmatrix}
\end{equation}
where $\pmb{\Phi}_{0, 0}$ indicates a single value of the basis spline for a given pixel and component in the $\phi$ dimension and
\begin{equation}
    \mathbf{R} = \begin{bmatrix}
    \mathbf{R}_{0, 0} & \mathbf{R}_{1, 0} & ... & \mathbf{R}_{j_r, 0} \\
    \mathbf{R}_{0, 1} & \mathbf{R}_{1, 1} & ... & \mathbf{R}_{j_r, 1} \\
    ... & ... & ... & ... \\
    \mathbf{R}_{0,\; n \times l} & \mathbf{R}_{1,\;n \times l} & ... & \mathbf{R}_{j_r,\; n \times l} \\
    \end{bmatrix}
\end{equation}
where $\mathbf{R}_{0, 0}$ indicates a single value of the basis spline for a given pixel and component in the $r$ dimension. We take the dot products of every column vector in $\pmb{\Phi}$ with every column vector in $\mathbf{R}$, and construct a new matrix
\begin{equation}
    \mathbf{X} =
    \left[
    \begin{bmatrix}
        \pmb{\Phi}_{0, 0}\mathbf{R}_{0, 0} \\
        \pmb{\Phi}_{0, 1}\mathbf{R}_{0, 1} \\
        ... \\
        \pmb{\Phi}_{0, n \times l}\mathbf{R}_{0, n \times l}
    \end{bmatrix},\;
    \begin{bmatrix}
        \pmb{\Phi}_{1, 0}\mathbf{R}_{0, 0} \\
        \pmb{\Phi}_{1, 1}\mathbf{R}_{0, 1} \\
        ... \\
        \pmb{\Phi}_{1, n \times l}\mathbf{R}_{0, n \times l}
    \end{bmatrix},\;...\;,
    \begin{bmatrix}
        \pmb{\Phi}_{j_\phi, 0}\mathbf{R}_{j_r, 0} \\
        \pmb{\Phi}_{j_\phi, 1}\mathbf{R}_{j_r, 1} \\
        ... \\
        \pmb{\Phi}_{j_\phi, n \times l}\mathbf{R}_{j_r, n \times l}
    \end{bmatrix}
    \right]
\end{equation}
which is the design matrix for our PRF model, which now has shape ($l \times n$, $j_\phi \times j_r$).

At the center of the PRF, there are naturally few sampled points. This is illustrated in Figure~\ref{fig:psfdemo1}, where there are fewer points at low values of $r$. Close to the center of the PRF, we do not have enough information to accurately fit for any $\phi$ dependence. As such, we do not build the cross terms for $r < 6"$ (equivalent to 1.5 \kepler pixels), and we only use the terms from $\mathbf{R}$. At this limit, we have no $\phi$ dependence. This means that while the tip of the PRF is allowed to vary as a function of $r$, there is no azimuthal dependence.

We fit to find the optimal weights $\mathbf{\hat{w}}$ for the design matrix $\mathbf{X}$, fitting to the log of flux in each pixel in the vector of the pixel values in the mean image $\mathbf{\bar{f}}$, normalized by an estimate of the true flux for each source in each pixel. Using Equations~\ref{eq:sigmainv0}, ~\ref{eq:what0} and ~\ref{eq:mod0} we fit

\begin{equation}
    \label{eq:y}
    \mathbf{y} = \log_{10}\Big(\frac{\mathbf{\bar{f}}}{\mathbf{\hat{f}}}\Big)
\end{equation}

\begin{equation}
    \label{eq:ye}
    \sigma_{\mathbf{y}} = \left|\frac{\sigma_{\mathbf{\bar{f}}}}{\frac{\mathbf{\bar{f}}}{\mathbf{\hat{f}}}\ln(10)}\right|
\end{equation}
where $\mathbf{\bar{f}}$ denotes the mean flux value of pixels across all time, and $\mathbf{\hat{f}}$ indicates the a priori source flux estimate in each pixel. $\sigma_{\mathbf{\bar{f}}}$ indicates the error on $\mathbf{\bar{f}}$. We firstly use \gaia flux estimate for source fluxes (denoted as $\mathbf{g}$ in the above sections) which provides our first estimate $\mathbf{\hat{f}_0}$, but we iterate and update this value as discussed below.



We use solve
\begin{equation}
    y = \mathbf{X}\cdot\mathbf{w}
\end{equation}

using Equations ~\ref{eq:sigmainv0} and ~\ref{eq:what0} to
find the best fitting weights $\mathbf{\hat{w}}$. These weights provide the best fit model to $\mathbf{y}$. We set the covariance of the data $(\mathbf{K_y})$ to a diagonal matrix, where the diagonal elements are the vector $\sigma_{\mathbf{f}}^2$.  Since we are working in log space, without including $\sigma_{\mathbf{y}}$ we would implicitly up-weight fainter pixels, i.e. the wings of the PRF.

We use $\pmb{\Lambda}$, a diagonal matrix, where the diagonal elements are set to a vector of 1's and 0's to set the weights of our fit. We build $\pmb{\Lambda}$ such that elements of $\mathbf{y}$ that are "contaminated" (contain flux from more than one source) are set to 0, and other elements are set to 1. This ensures that only uncontaminated pixels contribute to our fit\footnote{We note that while this procedure is mathematically correct, in practice if attempting to recreate this work (for example in Python), contaminated pixel should simply be masked arrays.}. We use this masking to produce a model of the PRF shape. When we ultimately fit this model to obtain the flux of each source, we will not use this mask, and will fit all pixels, including contaminated ones.

\subsection{Estimating the flux of every source}
\label{sec:findfluxvalues}
In this section, we attempt to find a vector of weights $\mathbf{v}$ containing the true flux values of every source in the image. The weights $\mathbf{\hat{w}}$ and design matrix $\mathbf{X}$ derived above can now be used to evaluate the "normalized" PRF shape

\begin{equation}
    \label{eq:shape_model}
    \mathbf{s} = 10^{\mathbf{X}\cdot\mathbf{\hat{w}}}
\end{equation}
where $\mathbf{s}$ is a vector describing the normalized PRF model in each pixel, with shape $l \times n$. To create the model of the scene, we then populate a sparse matrix $\mathbf{S}$ of shape ($l, n$) with the values of $\mathbf{s}$.
\begin{equation}
    \mathbf{S} = \textrm{mat}(\mathbf{s})
\end{equation}
where $\textrm{mat}$ indicates the linear transformation that converts a column vector into a matrix (the inverse of $\textrm{vec}$).

$\mathbf{S}$ is mostly made up of zeros (or extremely small values), and is valued only close to a source. $\mathbf{S}$ is now a matrix where, when dotted with some vector $\mathbf{v}$ of length $l$ (representing the intrinsic flux of each source in the image) we are able to build an estimate for the brightness of every pixel in the image, (i.e. the model of the scene)

\begin{equation}
    \label{eq:modelflux}
    \mathbf{\hat{f}} = \mathbf{S}\cdot\mathbf{v}
\end{equation}

where $\mathbf{\hat{f}}$ denotes our mean estimate of the image data. We can now fit to find estimates of the values $\mathbf{\hat{v}}$ by inverting Equation~\ref{eq:modelflux} and using Equations ~\ref{eq:sigmainv0} and ~\ref{eq:what0}. We find $\mathbf{\hat{v}}$ in a given image (note that the hat symbol denotes this as our best mean estimate of the true weights $\mathbf{v}$). This will yield an estimate of the flux value for every source. Note that we do not mask contaminated pixels when fitting for $\mathbf{\hat{v}}$, and so we naturally account for contamination.

In our framework we perform an iterative process, to better estimate the PRF shape and therefore the source flux. We fit the best fit shape model (Equation~\ref{eq:shape_model}), and then fit to find the best fitting PRF weights $\mathbf{\hat{v}}$ in the mean frame $\mathbf{\bar{f}}$. We evaluate $\mathbf{\hat{f}}$ (Equation~\ref{eq:modelflux}) at all pixels in the mean frame, and then mask pixels where the estimated flux would be less than $1e^-s^{-1}$ (where we assume we have reached the noise limit of the \kepler instrument.) We then update $\mathbf{y}$ with our updated estimate of the flux in each pixel ($\mathbf{g}$), update $\Lambda$ where necessary, and reject any significant outliers in our shape modeling. This enables us to iteratively remove "background" pixels, where there is not a significant contribution of flux. We perform this iteration three times. The final version of our model $\mathbf{\hat{f}}$ is shown in Figure ~\ref{fig:psfdemo1} for the mean frame. Though we start with a broad circular "aperture" to capture pixels that contain some flux, Figure ~\ref{fig:psfdemo1} demonstrates that we are able to use this iteration to remove irrelevant pixels, and evaluate the model only in the region directly related to the PRF.

Because we are modeling the entire scene in the image, $\mathbf{S}$ includes every source, and fits all sources simultaneously. This is powerful; it allows us to simultaneously fit sources that are are close to each other and overlapping, and still separate their true flux. Equation~\ref{eq:sigmainv0} also allows us to estimate the errors on $\mathbf{\hat{v}}$ (i.e. the flux errors of each source.) These errors naturally account for the increased uncertainty around contaminated sources.

The model described here assumes that all PRFs have the same shape. We know this to be incorrect. Inter- and intra-pixel sensitivity variations change the shape of the PRF depending on its location, but here we assume that these effects average out over a large enough number of sources. The PRF is also a function of the flux of the source. The \kepler detector is non-linear, (i.e. the recorded counts from a source is not linearly related to the flux of the source, see KIH.) Extremely bright sources (close to the saturation limit) also exhibit "halos" from internal reflections within the telescope. Here we have ignored the changes in PRF shape as a function of source brightness, but the model presented here could be updated to include this dependence.

\begin{figure}
    \centering
    \includegraphics[width=0.5\textwidth]{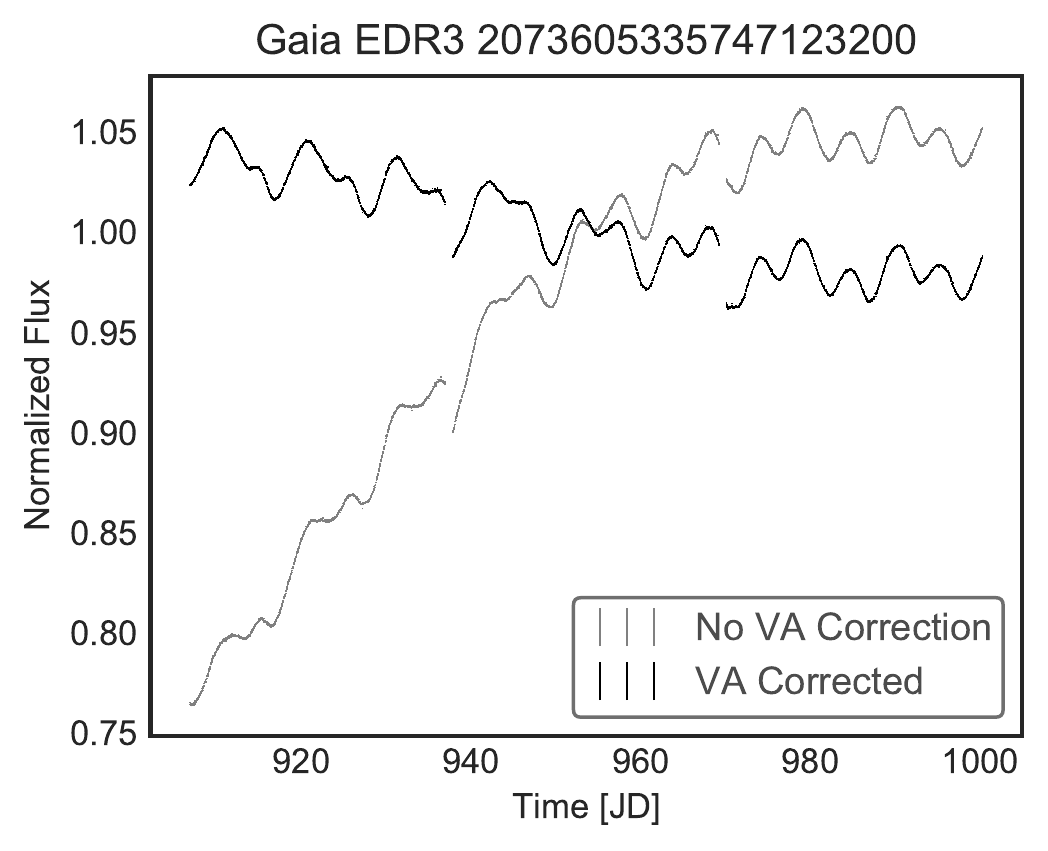}
    \caption{Demonstration of the model described in Sections \ref{sec:linearmodel} and ~\ref{sec:findfluxvalues}. Here a single bright source out of the \nsources test sources is shown. Plotted in grey is the derived flux of the source, with errors. The significant long term trend is due to velocity aberration, as the source slowly drifts approximately a pixel over the focal plane. The best fit flux when velocity aberration is accounted for (see Section~\ref{sec:motion}) is show in black. Some residual long term trends remain in the black light curve, suggesting that our model could be further improved. }
    \label{fig:demo_no_va}
\end{figure}

One of the key limitations in finding the flux of targets using Equations~\ref{eq:sigmainv0} and ~\ref{eq:what0} is that there is no way to make a non-Gaussian prior on $\mathbf{v}$. This means that in practice, sources are allowed to have negative flux values. For the most part, this is not a problem, but we find that there are some small numbers of cases of faint, crowded targets where source fluxes estimated with our method are negative, and so un-physical. In such cases, we can assume that our "scene" model is incorrect, and either the PRF model is incorrect, or the catalog of sources is incomplete or incorrect. In our tests these cases only occurred for faint, extremely crowded targets that fall off the edge of the detector. When iteratively fitting, we remove any targets from the model where the best fit flux ($\mathbf{\hat{v}}$) values go below zero, and refit, assuming there is no source at that location. It would also be possible to instead place narrow priors on these targets and enforce the flux must be close to the values reported by Gaia.

We can now find the best fitting flux value for every source as a function of time, $\mathbf{\hat{v}}_i$. For the $i$th image we fit $\mathbf{f}_i$, (the pixels in each $i$th frame), using Equations ~\ref{eq:sigmainv0} and ~\ref{eq:what0}, we find $\mathbf{\hat{v}}_i$, the vector of best fitting weights for each source, such that when dotted into $\mathbf{S}$ they provide the best fitting flux in every pixel (Equation~\ref{eq:modelflux}). We also find $\mathbf{K}_{\mathbf{v}_i}$, the covariance of the weights of each of the PRFs, which provides the errors on $\mathbf{v}_i$. We assume that the covariance of the data ($\mathbf{K_f}_i$) is well described by a simple diagonal matrix, where the diagonal terms are $\sigma_{\mathbf{f}_i}^2$.

Figure~\ref{fig:demo_no_va} shows a demonstration of the model in Equation~\ref{eq:modelflux} for every frame (grey line). (While our model fits all sources simultaneously, we show one source as a demonstration.) Figure~\ref{fig:demo_no_va} shows that there is a significant long term trend in the data. This trend is not astrophysical, and is instead due to velocity aberration (see KIH for further information), which causes the source to drift slowly over the duration of an observation. This drift is usually on the order of a single pixel. Because our model in Equation~\ref{eq:modelflux} is incapable of moving to account for this motion, the long term trend persists in our estimate of $\mathbf{v}_i$. For many science cases, such as searching for the short term transits of exoplanets, this long term trend is undesirable but could be removed simply (e.g. the \kepler Pipeline's CBV correction was designed to remove long term trends from velocity aberration, amongst other effects.) However, in cases where sources fall half on the detector (e.g. at the edge of TPFs) this drift will be degenerate with real, long term variability. As such, we will improve the model to account for "common" motion across sources in the scene.

The model we have built in this section is deliberately simplistic, and demonstrates how to address scene motion using a simple, low order polynomial. In some applications, it may be prudent to use more informative vectors to build $\mathbf{A}$. For example, including the \kepler CBVs into $\mathbf{A}$ may help address issues such as focus change, and using spacecraft position/centroid information may allow the model to account for roll motion in the \ktwo dataset. We leave this investigation as to the best way to construct $\mathbf{A}$ for \kepler to further work.

\subsection{Motion}
\label{sec:motion}
In our model we fit each cadence independently, despite the fact that each frame is highly correlated, and one frame will be highly predictive of the next. This is a reasonable approach; we want to obtain the flux of each source independently of the flux measured in neighbouring frames, because a priori we have no information about how a source should truly vary astrophysically. However, we would like to allow our model to take into account the motion of the scene for velocity aberration. To do this, we can update our model $\mathbf{S}$, and have the entire scene move in the same way that the data drifts. We will assume that the motion is small ($\lesssim$ the width of a pixel) and so the model at any given frame can be estimated as the model in the mean frame, multiplied by some small-valued function, describing the change in the model.

We will make the following implicit assumptions; 1) Kepler background is effectively zero, and can be ignored 2) sources, on average, do not vary astrophysically (instantaneous astrophysical flux for each source is usually very close to the average over time) 3) The motion is small, and varies smoothly over a long time period. In this section, we will use the same model for the PRF from Section~\ref{sec:linearmodel}, and update the way we fit for the flux of each source to account for some time dependence.

To update the model for velocity aberration, we first bin the data function of time, creating a smaller number of time averaged frames. We bin in this way for two reasons; firstly to increase signal to noise for the motion, and average over astrophysical noise, and secondly to decrease the amount of data that needs to be fit, and speed up our calculation of the model. To fit the drift, we divide the binned frames by the mean image $\mathbf{\bar{f}}$, normalizing each pixel time series to the average flux value. We will denote this binned, normalized flux as $\mathbf{f_b}$, is defined as

\begin{equation}
    \mathbf{f}_b = \mathbf{B}\cdot
    \left(
    \frac{\mathbf{f}}{\bar{\mathbf{f}}}
    \right)
\end{equation}
where
\begin{equation}
    \mathbf{B} =
    \begin{bmatrix}
    \frac{1}{b}...\frac{1}{b} & & & \\
    & \frac{1}{b}...\frac{1}{b} & &\\
    & & \frac{1}{b}...\frac{1}{b} &\\
    & & & ...
    \end{bmatrix}
\end{equation}
where $b$ is the number of cadences we sum to create each bin (in our example, we use $b$=200 which provides a resolution of ~4 days and is adequate to capture \kepler's long term motion). The matrix $\mathbf{B}$ is a made up of entries of $\frac{1}{b}$ for cadences we wish to bin. $f_b$ is now a matrix, with shape ($l \times n$, $m / b$) Assuming most sources on average do not vary, $\mathbf{f_b}$ should only vary due to instrumental effects, and velocity aberration.

To fit common motion in the scene we will use another basis spline. In this case, unlike when fitting the PRF shape, there is no radial symmetry, the entire image drifts slowly in the $x$ and $y$ dimension. As such, we will fit a basis spline in Cartesian coordinates.

Using the same methods discussed in Section~\ref{sec:linearmodel}, we build a simple 2D basis spline design matrix in each Cartesian dimension, $\pmb{\delta X}$ and $\pmb{\delta Y}$, and calculate all the cross terms to build a design matrix, which we will denote as $\mathbf{A_0}$.

\begin{equation}
    \pmb{\delta X} = \begin{bmatrix}
    \pmb{\delta X}_{0, 0} & \pmb{\delta X}_{1, 0} & ... & \pmb{\delta X}_{j_x, 0} \\
    \pmb{\delta X}_{0, 1} & \pmb{\delta X}_{1, 1} & ... & \pmb{\delta X}_{j_x, 1} \\
    ... & ... & ... & ... \\
    \pmb{\delta X}_{0,\; c} & \pmb{\delta X}_{1,\;c} & ... & \pmb{\delta X}_{j_x,\;c} \\
    \end{bmatrix}
\end{equation}
where $\pmb{\delta X}_{0, 0}$ indicates a single value of the basis spline for the first pixel and first component in the $\delta x$ dimension and
\begin{equation}
    \mathbf{\delta Y} = \begin{bmatrix}
    \pmb{\delta Y}_{0, 0} & \pmb{\delta Y}_{1, 0} & ... & \pmb{\delta Y}_{j_y, 0} \\
    \pmb{\delta Y}_{0, 1} & \pmb{\delta Y}_{1, 1} & ... & \pmb{\delta Y}_{j_y, 1} \\
    ... & ... & ... & ... \\
    \pmb{\delta Y}_{0,\; c} & \pmb{\delta Y}_{1,\;c} & ... & \pmb{\delta Y}_{j_y,\; c} \\
    \end{bmatrix}
\end{equation}
where $\pmb{\delta Y}_{0, 0}$ indicates a single value of the basis spline for the first pixel and first component in the $\delta y$ dimension. $c$ is the length of each column vector
\begin{equation}
    c = l \times n \times m/b,
\end{equation}
 and $j_x$ and $j_y$ are the number of basis spline components (column vectors) in $\pmb{\delta X}$ and $\pmb{\delta Y}$ respectively. We take the dot products of every column vector in $\pmb{\delta X}$ with every column vector in $\pmb{\delta Y}$, and construct a new matrix
\begin{equation}
    \mathbf{A}_0 =
    \left[
    \begin{bmatrix}
        \pmb{\delta X}_{0, 0}\pmb{\delta Y}_{0, 0} \\
        \pmb{\delta X}_{0, 1}\pmb{\delta Y}_{0, 1} \\
        ... \\
        \pmb{\delta X}_{0, n \times l}\pmb{\delta Y}_{0, c}
    \end{bmatrix},\;
    \begin{bmatrix}
        \pmb{\delta X}_{1, 0}\pmb{\delta Y}_{0, 0} \\
        \pmb{\delta X}_{1, 1}\pmb{\delta Y}_{0, 1} \\
        ... \\
        \pmb{\delta X}_{1, n \times l}\pmb{\delta Y}_{0, c}
    \end{bmatrix},\;...\;,
    \begin{bmatrix}
        \pmb{\delta X}_{j_x, 0}\pmb{\delta Y}_{j_y, 0} \\
        \pmb{\delta X}_{j_x, 1}\pmb{\delta Y}_{j_y, 1} \\
        ... \\
        \pmb{\delta X}_{j_x,c}\pmb{\delta Y}_{j_y, c}
    \end{bmatrix}
    \right]
\end{equation}
which is the design matrix for our time dependent model, which now has shape ($l \times n \times m/b$, $j_x \times j_y$). We then create a time dependent design matrix $\mathbf{A}$ by taking the element-wise product of $\mathbf{A}_0$ and a 3rd order polynomial in time $\mathbf{t}$, assuming that this is a reasonable model for the slow, long-term velocity aberration. This gives us a design matrix as a function of time
\begin{equation}
    \label{eq:A}
    \mathbf{A} =
    \begin{bmatrix}
        \mathbf{A}_0, & \mathbf{A}_0 \circ \mathbf{t}_b, & \mathbf{A}_0 \circ \mathbf{t}_b^2, & \mathbf{A}_0 \circ \mathbf{t}_b^3
    \end{bmatrix}
\end{equation}
which has shape ($l \times n \times m/b$, $\times j_x \times j_y \times$ 4). $\mathbf{t}_b$ indicates a vector of the binned time values (corresponding to the average time at every element of $\mathbf{f}_b$) with length $l \times n \times m/b$. $\circ $ indicates the element-wise product.

We then fit the model to the normalised, binned data $\textrm{vec}(\mathbf{f}_b)$ using Equations ~\ref{eq:sigmainv0} and ~\ref{eq:what0} to find $\mathbf{\hat{u}}$, the best fitting weights of the time dependent matrix $\mathbf{A}$.

The motion model at the $i$th frame is then simply
\begin{equation}
    \mathbf{l}_{i} =
    \begin{bmatrix}
    \mathbf{A}_0, &
    \mathbf{A}_0 t_i, &
    \mathbf{A}_0 t_i^2, &
    \mathbf{A}_0 t_i^3
    \end{bmatrix}
    .\mathbf{\hat{u}}
\end{equation}
where $t_i$ indicates the time at the $i$th frame. $\mathbf{l}_{i}$ is a vector whose entries scale the mean PRF shape in a given frame ($\mathbf{s}_i$) to the correct shape given the motion in each frame. We iteratively fit this model to the normalised $\mathbf{f_b}$, masking out sources that show a significant discrepancy from the motion model (suggesting that there is large scale astrophysical variability).

To build the PRF shape model matrix $\mathbf{S}_i$ that accounts for this motion in the $i$th frame, we simply populate the matrix with the element-wise multiplication of $\mathbf{s}$ and $\mathbf{l}_i$

\begin{equation}
    \mathbf{S}_{i} = \textrm{mat}(\mathbf{s}\cdot\mathbf{l}_{i})
\end{equation}
We then fit to find the best fitting weights $\mathbf{v}_i$ for each source, in each frame, using each unique $\mathbf{S_i}$ model of the scene. We derive the flux for every source, simultaneously, accounting for common motion in the scene.

\begin{figure}
    \centering
    \includegraphics[width=0.5\textwidth]{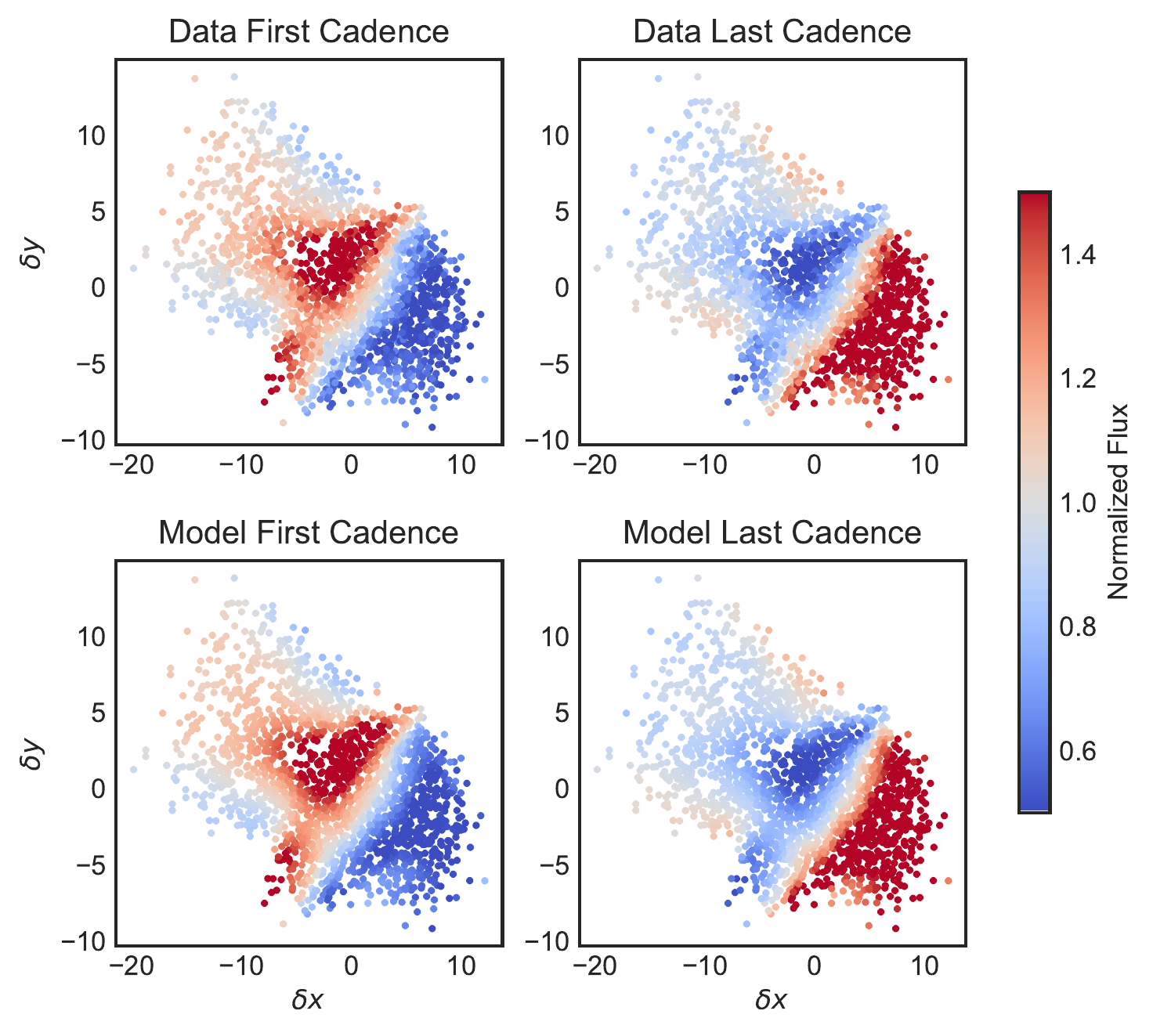}
    \caption{Example of our model for long term motion in our test dataset. \emph{Top}: The flux in of each source, when mean normalized ($\frac{\pmb{f_b}}{\pmb{\bar{f}}}$). \emph{Bottom}: The model described in Section~\ref{sec:motion}.
    Blue points indicate flux is less than one, and red points indicate flux is greater than one. The effect of velocity aberration is clear, where sources drift from the top left corner to the bottom right corner as a function of time. This figure also captures the slight changes in PSF shape due to temperature changes and focus changes. Our model shows good agreement with the data.}
    \label{fig:motion}
\end{figure}

Figure~\ref{fig:motion} shows the normalized, binned data $\mathbf{f}_b$ as a function of the $\delta x$ and $\delta y$ dimensions, alongside our best fitting model. Figure~\ref{fig:motion} shows that in the case of our demonstration data, sources are on average brighter in the top left corner of $\delta x$, $\delta y$ space at the start of the image stack, and brighter in the bottom right corner towards the end of the image stack (upper panels). This shows the sources are drifting, over time, from one part of $\delta x$, $\delta y$ space to another. Our model (lower panels) accounts for this motion.

Figure~\ref{fig:demo_no_va} shows the results of fitting the model, including velocity aberration, in black. The long term drift due to velocity aberration has been largely removed, leaving the astrophysical variability intact. While this has removed common motion from all sources, this method can not account for local effects such as a change in sensitivity of the pixels. Spatially varying intra-pixel sensitivity is a known effect in \kepler, which causes \ktwo systematic noise as the telescope roll motion moves the source over differently sensitive parts of the pixel. As such, while this improves the photometry, it can not completely account for this motion. Additionally, a more complete model would account for differential velocity aberration across the image. We find this method is adequate to improve the photometry of a source that is half off the edge of the detector.

\subsubsection{Focus change}

As discussed in the \cite{kih}, the \kepler instrument underwent significant focus change during observations, in particular during the start of data collection (or after any data collection gaps) as the telescope changed temperature. This focus change causes a significant shape change to the PRF, and in some cases a significant source drift. Focus change can be seen in Figure~\ref{fig:results} after gaps in data collection, as a significant "hook" shaped systematic. In this work, we do not account for this focus change. Instrumental systematics such as this are well corrected by methods such as the \kepler pipeline's CBV correction, and so we assume that focus change can be corrected in post-processing steps. It would be possible to create a more flexible time dependent model, and use more time points (i.e. $b<200$) to account for some of this short term PRF shape change.

\subsection{Sparsity}

In the above sections, we have discussed a model that fits the scene in a given stack of \kepler images. For our test set, we are able to fit the PRF shape model and motion model, and produce light curves of all \nsources sources in under 4 minutes running on a personal laptop\footnote{\label{note1} using 2017 MacbookPro, 16GB Ram, 2.9 GHz Intel Core i7 processor}. This speed is only possible when we take advantage of the sparsity of the matrices described above. Many of the variables discussed above are extremely sparse, as pixels far from sources effectively contribute zero flux. By using \texttt{scipy}'s \texttt{sparse} library \citep{scipy}, we are able to capitalize on the sparsity of these matrices, and avoid loading large arrays of zeros into memory, making it tractable to fit the scene for this data set on an average machine. The tools we have developed for this work are available in an open source Python tool \footnote{\url{https://github.com/SSDataLab/psfmachine}} under an MIT License, and v1.0.0 is archived in Zenodo \citep{PSFMachine}. This process is expandable to larger datasets; for example it is possible to estimate the PRF model and source flux of every source in a \kepler Full Frame Image. For a single channel, this requires building a matrix of $\approx$100,000 pixels, and $\approx$ 13,250 sources. Solving to find the PRF model and the weights of the sources then takes approximately 150 seconds for a single frame\footnoteref{note1}, and requires 500Mb of memory allocation. This demonstration shows that \acnm is scalable, and applicable to larger datasets (for example, the \tess dataset).

\subsection{Diagram of the method}
\label{sec:flowchart}

\begin{figure}
    \centering
    \includegraphics[width=0.5\textwidth]{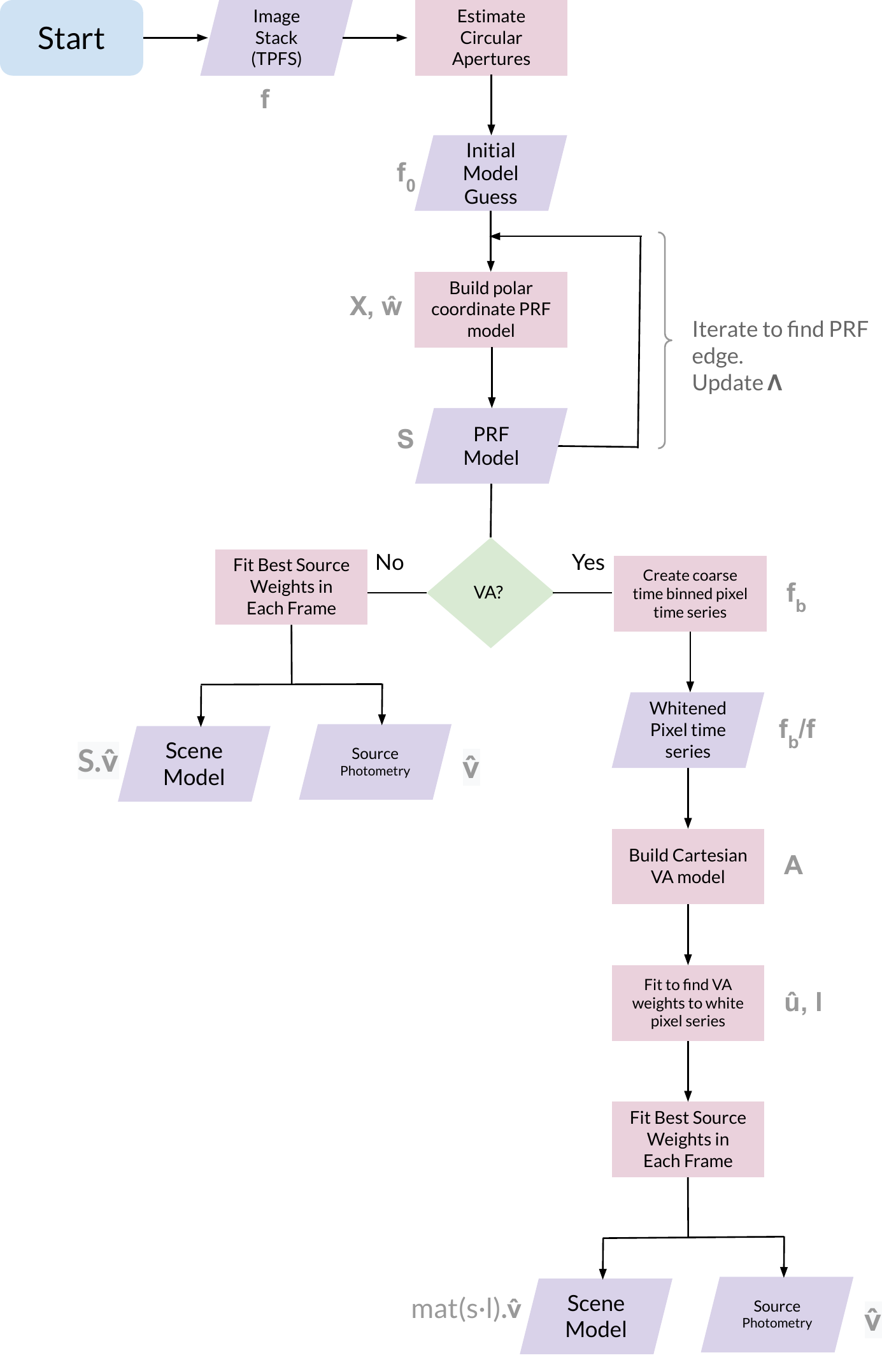}
    \caption{Sketch of the flow of the \acnm method described in this work. Here VA stands for velocity aberration, which we assume is the largest source of long term time trends in the input data. The quantities described in this work related to each part of the proceedure are written in grey next to each stage. A full description of all the variables used in this work is given in Table~\ref{tab:nomenclature} in the Appendix.}
    \label{fig:flowchart}
\end{figure}

Figure ~\ref{fig:flowchart} shows an overview of the key stages involved in our method. Depending on the specific requirements of a given dataset, users may either choose to fit with or without accounting for long term trends. By following Sections~\ref{sec:edge}, ~\ref{sec:linearmodel}, and ~\ref{sec:findfluxvalues} users can build simple photometry of sources (left side of flow chart in Figure ~\ref{fig:flowchart}), and by following Sections~\ref{sec:edge}, \ref{sec:linearmodel} and ~\ref{sec:motion} users can build photometry of sources, accounting for long term motion (right side of flow chart in Figure ~\ref{fig:flowchart}).

\subsection{Results}
\label{sec:results}

\begin{figure*}
    \centering
    \includegraphics[width=\textwidth]{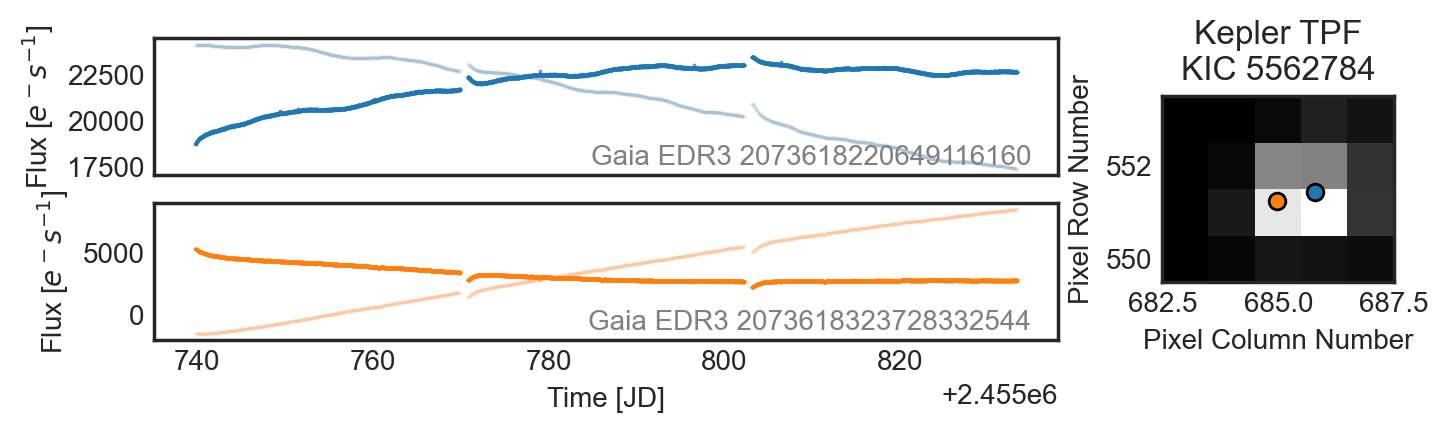}
    \caption{Results of our model fit for \koi. \emph{Right:} Single frame of the TPF data for \koi. Orange and blue points indicate the two sources that are in the data, as identified by \gaia. \emph{Left:} The best fit light curves from our model of the scene for the two sources in the TPF. These sources are separated by approximately one pixel, and the fainter (orange) source is approximately 2 magnitudes fainter. Fainter lines indicate the best fit light curve without accounting for motion. The light curves still retain some long term trends due to motion, but are improved by our treatment of velocity aberration.}
    \label{fig:results}
\end{figure*}

\begin{figure}
    \centering
    \includegraphics[width=0.5\textwidth]{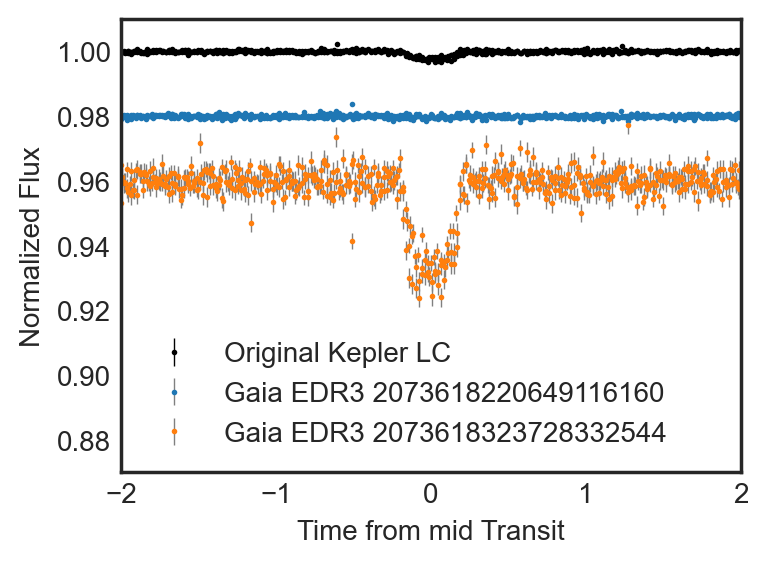}
    \caption{The light curves from Figure~\ref{fig:results}, folded at the best fit period for \koi and binned. Each light curve is separated by an offset for clarity. Shown in black is the original \kepler data from the Pipeline. Aperture photometry causes the signals of the two sources to be mixed. In orange is the light curve for the true host of the transiting signal, and in blue is the light curve of the contaminant. Our method is able to 1) well separate both targets, despite them being close on the detector and the contaminant being much brighter and 2) provide realistic errors, that account for the uncertainty arising from the contaminated source.}
    \label{fig:results_planet}
\end{figure}

In Sections~\ref{sec:linearmodel}, ~\ref{sec:findfluxvalues} and ~\ref{sec:motion} we have discussed our method for modeling the scene in an stack of astronomical images, solving for the flux of all sources. Our method derives photometric light curves of each of the \nsources sources in the scene, with errors. \acnm is able to fit the instrument PRF and velocity aberration, and then model the flux in each source as a function of time in under 4 minutes on a personal computer. In this section we will discuss these results for the target of interest; \koi.

Figure ~\ref{fig:results} shows the \kepler TPF data for \koi, and the best fit light curves for both targets in the data. Figure ~\ref{fig:results_planet} shows the same light curves, alongside the \kepler pipeline light curve for \koi, flattened with a simple Savitsky Golay filter, folded at the period of \koi, and binned. The transiting signal clearly originates from \gsourcea, and our method is able to separate both targets exceptionally well.

The transiting signal around \gsourcea caused \koi to be identified as a false positive exoplanet candidate by the \kepler mission. It was discounted due to a centroid offset during transit. In this case, the transiting signal still remains a viable case for a transiting planet, with a depth potentially consistent with a very large planet or sub-stellar companion, and a transit like shape. 

\gsourcea does not have a luminosity or radius value specified in \gaia EDR3\footnote{\gaia EDR3 does not provide updated stellar parameters over previous data releases, and so we use \gaia DR2 here.}, due to the large recorded parallax error. Using the relationships from the Gaia Data Release Documentation \citep[see Section 14.1.1 in ][]{gaiadr2doc}, we are able to create an estimate of the radius and luminosity of the target, given \gaia EDR3 estimates of the effective temperature and line-of-sight extinction A$_g$, and \gaia DR2 estimates of the mean $g$ magnitude and parallax. \gaia provides the information required to build Gaussian priors on each of these parameters. We use a simple luminosity power law for stellar mass, and allow the power ($\alpha$) parameter to vary uniformly between 3 and 4. This enables us to vary the stellar density, with a reasonable physical relationship with luminosity. We model the light curve for \gsourcea using the \texttt{exoplanet} tool \citep{exoplanet:exoplanet} and \texttt{pymc3} \citep{exoplanet:pymc3}, which enables us to set up Gaussian priors based on the \gaia data, and fit for limb darkening.

The results of our transit model are shown in Table~\ref{tab:parameters}. \gaia DR3 quotes a parallax value for \gsourcea of 0.342$\pm0.267$ $mas$. Since the parallax errors for this target are so large, the luminosity, mass and radius errors are also broad. Despite these broad errors in stellar parameters, we obtain a radius estimate for the transiting object of $2.52\pm_{0.14}^{0.16}$R$_{jup}$. This radius is consistent with a small sub-stellar object around \gsourcea. This could imply that either the transit is from a very large planetary companion, or a small sub-stellar or stellar companion.

\begin{table}
\begin{tabular}{lll}
\toprule
         Parameter &                           Description &                          Value \\
\midrule
     R$_*$\dotfill &            Radius [$R_\odot$]\dotfill &      1.72 $\pm_{-0.09}^{0.09}$ \\
     M$_*$\dotfill &              Mass [$M_\odot$]\dotfill &      1.16 $\pm_{-0.07}^{0.06}$ \\
 T$_{eff}$\dotfill &     Effective Temperature [K]\dotfill &  4990.0 $\pm_{-170.0}^{160.0}$ \\
     u$_1$\dotfill &  Limb Darkening Coefficient 1\dotfill &        0.58 $\pm_{-0.2}^{0.2}$ \\
     u$_2$\dotfill &  Limb Darkening Coefficient 2\dotfill &      0.07 $\pm_{-0.33}^{0.33}$ \\
\midrule
    R$_P$\dotfill &           Radius [$R_{jup}$]\dotfill &              2.52 $\pm_{-0.14}^{0.16}$ \\
        P\dotfill &                Period [days]\dotfill &          25.337 $\pm_{-7e-05}^{7e-05}$ \\
    T$_0$\dotfill &       Transit Mid Point [JD]\dotfill &  2455025.9156 $\pm_{-0.0011}^{0.0011}$ \\
        i\dotfill &       Inclination [$^\circ$]\dotfill &              89.45 $\pm_{-0.44}^{0.4}$ \\
      $a$\dotfill &         Semi-Major Axis [AU]\dotfill &             0.306 $\pm_{-0.02}^{0.02}$ \\
  $a/R_*$\dotfill &      Semi-Major Axis / R$_*$\dotfill &                38.1 $\pm_{-0.7}^{0.7}$ \\
 T$_{14}$\dotfill &             Duration [hours]\dotfill &                21.5 $\pm_{-0.7}^{0.9}$ \\
 T$_{eq}$\dotfill &  Equilibrium Temperature [K]\dotfill &             571.0 $\pm_{-14.0}^{14.0}$ \\
 \midrule
\dotfill&RA\dotfill&298.87096 $\pm$ 0.00021\\
\dotfill&Dec\dotfill&40.72015 $\pm$ 0.00024\\
\dotfill&Proper Motion RA [mas/yr]\dotfill&-4.85 $\pm$ 0.27\\
\dotfill&Proper Motion Dec [mas/yr]\dotfill&-4.21 $\pm$ 0.32\\
\dotfill&Parallax [mas/yr]\dotfill&0.34 $\pm$ 0.27\\
\bottomrule
\end{tabular}

\caption{Best Fit parameters the transiting signal around \gsourcea}
\label{tab:parameters}
\end{table}



\subsubsection{Comparison with PDCSAP Flux}

\begin{figure}
    \centering
    \includegraphics[width=0.5\textwidth]{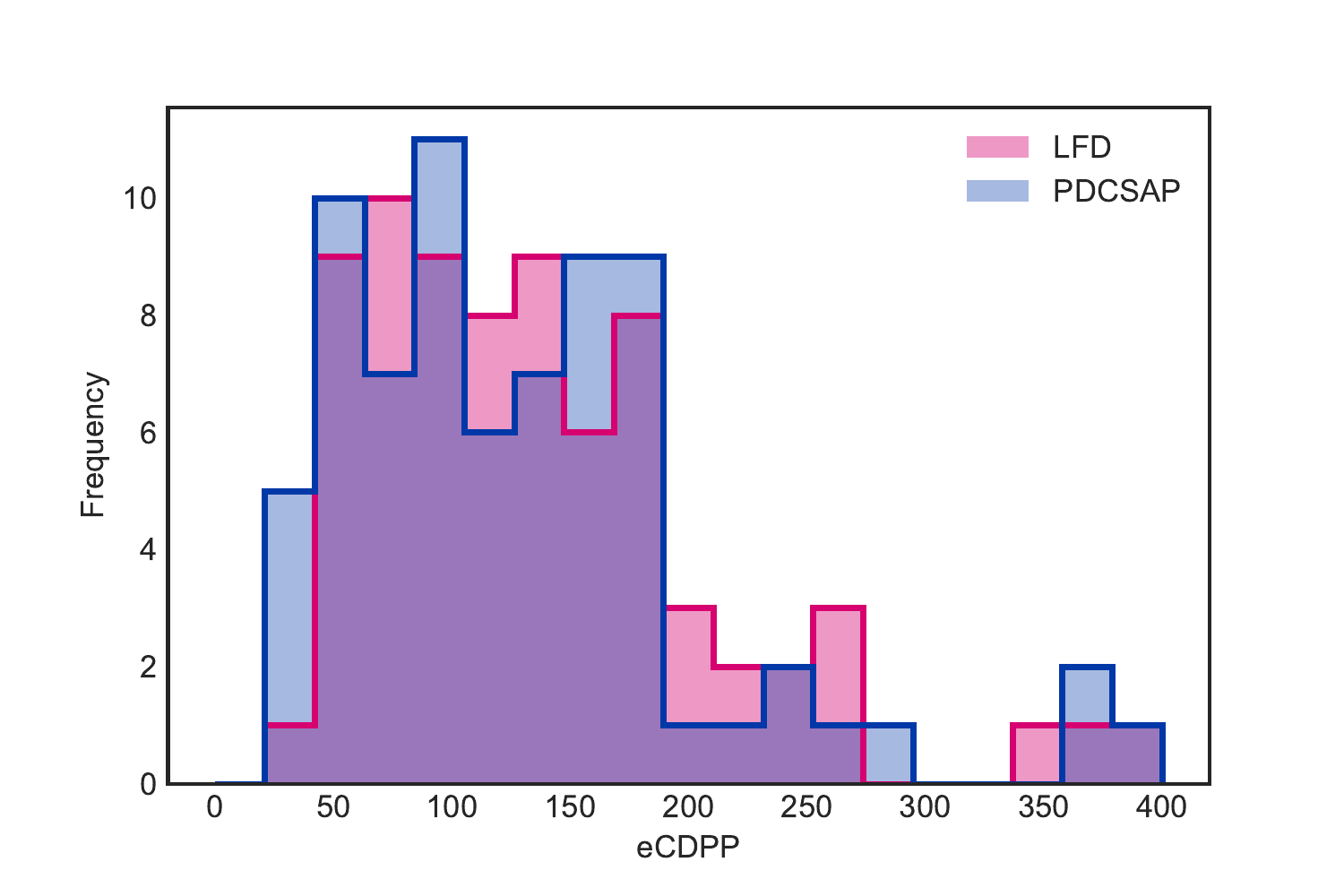}
    \caption{Comparison of the CDPP metric estimated by \lk for the original \kepler PDCSAP light curves, and the light curves produced in this work. CDPP is given in units of parts per million. Lower CDPP values indicate lower photometric noise in the light curve. We find we are able to achieve precision consistent with the pipeline products. (In this comparison we only compare targets that are not crowded, as our method separates crowded targets into individual light curves.)}
    \label{fig:pdcsap_comp}
\end{figure}

The \texttt{lightkurve} Python package provides the ability to calculate an estimated Combined Differential Photometric Precision metric, similar to the metric used by the \kepler Pipeline to determine light curve quality. \lk's \texttt{estimate\_cdpp} function use the simpler “sgCDPP proxy algorithm” discussed by \cite{gilland} and \cite{vancleve}. This single numeric value can be used as a measure of the noise properties of the data. We compare the light curves obtained from \acnm with the PDCSAP light curves available for targets in our sample. We remove any targets where there is significant crowding, as our method separates crowded targets into individual light curves. Figure~\ref{fig:pdcsap_comp} shows the estimated CDPP for both sets. We find that in our test case \acnm is able to reach similar precision to the PDCSAP products, and does not significantly increase or decrease the photometric noise of the time series compared to PDCSAP.

\subsubsection{Validation}

The methods described here have been demonstrated using \koi, but would be equally applicable to other Kepler Objects of Interest that were discounted due to a significant centroid shift, to search for further planetary candidates. In literature, planetary candidates are usually validated as true planets using a suite of different approaches \citep[for example the Kepler Robovetter][]{robovetter} Many of these approaches are invalid in the case of highly contaminated targets whos photometry is extracted in this way. Tests such as centroiding tests, are not valid given our approach, (centroiding tests are only applicable in the case of aperture photometry). Futher photometric observations at high spatial resolution are frequently employed to search for contaminating objects, but in this case, we are aware of a significant contaminant next to the object in question. Ground-based or space-based follow-up to re-observe the transits would require high signal to noise and spatial resolution to be able to separate these blended targets, and resolve the small transit around the faint background target. These factors would make it more difficult to confirm transiting signals from KOIs analyzed in this way, but further observations with high resolution imaging, high resolution spectra, and/or large ground based facilities, may be able to confirm targets similar to the transiting signal around \gsourcea.

\section{Discussion}
\label{sec:discussion}

In this work we have presented a framework for modeling a field of stars from \kepler, and demonstrated the power of the method to separate even highly contaminated sources. In this section we will discuss some of the limitations and assumptions in our framework.

Our model is built based on an input catalog of source positions, built from the \gaia EDR3 catalogue, with proper motions accounted for. If this input catalogue is missing sources, our model is not able to account for this, and our results will be inaccurate. We find this assumption to be valid for \kepler, where the pixel scale is much larger than the spatial resolution of \gaia. In this work, we have only accounted for sources in the \gaia EDR3 catalogue down to 18th magnitude in \gaia $g$, and have assumed that "missing" sources fainter than 18th magnitude will not adversely affect our results. This assumption could be relaxed in future work.

We have also assumed that there is no distortion over the image, and PSF shapes do not vary. For \kepler and \tess data, this assumptions hold for reasonably small patches of the CCD (e.g. 100s of pixels), however the PSF shape is known to distort towards the edges of the focal plane (e.g. see KIH, or the TESS Instrument Handbook). The model presented in this work could be updated to allow the PRF shape model (Equation ~\ref{eq:shape_model}) to vary as a function of position on the detector, as well as $\phi$ and $r$, and so it is possible to relax this assumption. Similarly, the model could be updated to have a dependence on source flux, to account for other pixel-level effects in the detector, for example non-linearity.

In this work we have assumed that the PRF shape does not change over time. This is known to be false.

In the case of \kepler, the PRF is known to change shape over time due to several effects.
Focus change, due to changing temperatures in the spacecraft, is known to cause changes in the PSF position and shape on the detector \citep[see KIH][]{kih}.
The PRF shape is also known to be altered by the source color \citep[see][]{chromo}.
Focus change is the largest magnitude effect, and similar approaches to those undertake in Section~\ref{sec:motion} could be used to mitigate the effect of focus change by updating $\mathbf{A}$ to include information on rapid focus change at cadences close to the start of data collection.
The effect of chromatic aberration is demonstrated to be extremely small, and unlikely to cause a significant adverse effect on the light curves derived by this method. However, it would be possible to analyse the residuals between the model built in this work and the data, and search for significant deviations from the model that can be attributed to a shape change due to chromatic aberration (i.e. a source changing color).

For \tess the PRF shape can change significantly frame to frame due to intra-cadence motion caused by "jitter". In our method we assume that these changes to the PRF shape are small, and common to all sources. We assume that these PRF shape changes cause systematics in the retrieved time-series that can then be corrected by post-processing, (which is the approach used for simple aperture photometry). In our method we create a PRF model for the average frame, but it would also be possible build a unique PRF shape model at every frame. In practice, we find this produces worse photometry, since estimating the PRF shape using a single frame provides a noisy estimate compared with the mean frame. For cases where there is a significant shape change, and a large number of sources (e.g. perhaps large ground-based surveys affected by seeing), our assumptions may not hold.

We have assumed that the bulk motion in the scene is 1) $\lesssim$ than the width of the PRF 2) smooth over time 3) common across all sources. We require the motion to be less than the width of the PRF for our approach to be valid (Section~\ref{sec:motion}), since we essentially assume that the PRF in any given frame can be modeled as the PRF in the mean frame, multiplied by some (small) factor. For large motion, this does not hold. In \kepler data, and largely in \tess these three assumptions about motion are true. For \ktwo data, motion is small, and common across sources, but the 6-hour roll motion breaks the assumption that motion is smooth over time. We suggest this could potentially be alleviated by changing the time dependence in Equation~\ref{eq:A} from a simple polynomial to a function of the centroid position, similar to the approach in \cite{vanderburg}.

Finally, we have largely assumed that the interpixel and intrapixel sensitivity variations of the instrument is negligible in this work. For interpixel sensitivity variations, we assume the \kepler Pipeline is able to accurately flat-field the data. However, the \kepler detector is known to exhibit significant intrapixel sensitivity variations. It is these variations that, when coupled with the significant roll motion of the \kepler spacecraft, cause the distinct 'sawtooth' noise pattern in \ktwo data. When we build our PRF model, we are implicitly assuming that over a large number of sources, these variations average out, so that we can fit a smooth function to the PRF with the average intrapixel sensitivity. When we fit for the weights of a source, accounting for scene motion, we are not accounting for this sensitivity change across the pixel in the method described in this work. For work with \kepler, this may cause a long term trend in the data that is not accounted for by our velocity aberration correction (e.g. see Figure~\ref{fig:demo_no_va}). For \kepler, and for the purpose of finding short period variability, our approach is reasonable. For \ktwo data, an additional correction for intrapixel sensitivity will be needed. Further investigation is needed to find the best approach to including intrapixel sensitivity in the model described here.

\section{Summary}

In this work we have presented a method of obtaining photometry we name \acronym, where we estimate a model for the Pixel Response Function of sources, and then model the flux of every source simultaneously, fixing the source positions to those estimated by \gaia. \acnm is general and fast. The method we have presented in this work is highly applicable to crowded sources, as we have demonstrated with \koi. We are able to produce accurate photometry of sources that are separated by ~1 pixel on the detector, with a contrast of 2 magnitudes. \acnm is capable of fitting the flux of sources even when significant portions of the PRF fall off the detector, by fitting the wings of the PRF.

A further investigation is needed to full understand the limits of \acnm in crowded fields, both as a function of target separation and target magnitude in crowded regions. At some spatial separation, and contrast, our model will not be able to adequately separate targets. We leave this investigation to future work. Such an investigation may help us develop methods and practices for vetting planets identified in \acnm light curves.

The \kepler and \ktwo super stamp data, which cover several clusters, may benefit considerably from analysis using this approach. The photometry of sources in the \ktwo microlensing super stamp, which has many hundreds of thousands of sources, could be extracted using this method. However, we rely on a complete catalog for our method, and so for fields towards the galactic center could be difficult to analyse, since \gaia may be less complete.

NASA's \tess mission has much larger pixels (27"), a broader PRF and a fainter magnitude limit than \kepler, resulting in much more significant crowding. We suggest that \acnm is highly applicable to \tess fields, and will be useful for separating contaminants from exoplanet host stars, to better estimate the radius of transiting objects in the \tess survey.

\section{Acknowledgements}
This paper includes data collected by the Kepler mission and obtained from the MAST data archive at the Space Telescope Science Institute (STScI). Funding for the Kepler mission is provided by the NASA Science Mission Directorate. STScI is operated by the Association of Universities for Research in Astronomy, Inc., under NASA contract NAS 5–26555. This work has made use of data from the European Space Agency (ESA) mission
{\it Gaia} (\url{https://www.cosmos.esa.int/gaia}), processed by the {\it Gaia}
Data Processing and Analysis Consortium (DPAC,
\url{https://www.cosmos.esa.int/web/gaia/dpac/consortium}). Funding for the DPAC
has been provided by national institutions, in particular the institutions
participating in the {\it Gaia} Multilateral Agreement. This research made use of \textsf{exoplanet} \citep{exoplanet:exoplanet} and its
dependencies \citep{exoplanet:agol20, exoplanet:arviz, exoplanet:astropy13,
exoplanet:astropy18, exoplanet:exoplanet, exoplanet:kipping13,
exoplanet:luger18, exoplanet:pymc3, exoplanet:theano}.
Funding for this work for CH and JMP is provided by grant number 80NSSC20K0874, through NASA ROSES.

\appendix
\section{Paper Naming Conventions}
\label{sec:table}
In this paper we use lower-case italics to describe single valued variables, lower-case bold to describe vectors and upper case bold to describe matrices.
\begin{table}[H]
    \centering
    \begin{tabular}{c|p{15cm}}
      \textbf{Parameter}   &  \textbf{Description} \\
        $n$ & Number of pixels\\
        $m$ & Number of cadences\\
        $l$ & Number of sources\\
        $i$ & Indicates the $i$th frame in a stack of images\\
        $\sigma_y$ & Indicates the measurement errors on a variable (in this case a dummy variable y)\\
        $\mathbf{K}_\mathbf{y}$ & Indicates the covariance matrix of a vector (in this case a dummy vector $\mathbf{y}$)\\
        $\hat{y}$ & We use the "hat" symbol to denote a mean estimate of a variable, (in this case a dummy variable $y$)\\
        $\bar{y}$ & We use the "bar" symbol to denote the mean of a variable in time, (in this case a dummy variable $y$)\\
        $x$, $y$ & Cartesian coordinates\\
        $r$, $\phi$ & Polar coordinates\\
$\pmb{\delta x}$ & Vector of separations from a source in Cartesian coordinate $x$, with length $l \times n$\\
        $\pmb{\delta y}$ & Vector of separations from a source in Cartesian coordinate $y$, with length $l \times n$\\
        $\mathbf{r}$ & Vector of separations from a source in polar coordinate $r$, with length $l \times n$\\
        $\pmb{\phi}$ & Vector of separations from a source in polar coordinate $\phi$, with length $l \times n$\\
        $\mathbf{f}$ & Vector of pixel fluxes with length $n \times m$ (the data)\\
        $\mathbf{\bar{f}}$ & Vector of pixel fluxes averaged over time with length $n$\\
        $\mathbf{g}$ & Vector of pixel flux estimates built from source flux estimates from Gaia EDR3, with length $l \times n$\\
        $\mathbf{X'}$ & Design matrix for circular apertures, with shape ($l \times n$, 9) \\
        $\mathbf{\hat{w_0}}$ & Best fitting weights for design matrix $\mathbf{X'}$ to model circular apertures, vector with length 9 \\
        $\mathbf{\hat{f}_0}$ & Estimate of the flux in each pixel from the initial, circular model, vector with length $n$\\
        $\mathbf{\hat{f}}$ & Vector of pixel flux estimates with length $n$\\
        $\mathbf{\hat{v}}$ & Vector of source flux weights, with length $l$\\
        $\mathbf{s}$ & Vector of the normalized, PRF shape model with length $l \times n$.\\
        $\mathbf{S}$ & Matrix of the normalized, PRF shape, with shape ($l$, $n$), this is the matrix version of $s$, and when dotted with $\mathbf{\hat{v}}$ will give the scene model.\\
        $\Lambda$ & Matrix of "weights" for each pixel, which is a diagonal matrix consisting of 1's for uncontaminated pixels, and 0's for contaminated pixels, has shape ($n$, $n$)\\
        $\pmb{\Phi}$ & Design matrix of basis spline components in the $\phi$ dimension, has shape ($l \times n$, $j_\phi$)\\
        $\mathbf{R}$ & Design matrix of basis spline components in the $r$ dimension, has shape ($l \times n$, $j_r$)\\
        $j_\phi$ & Number of components in the design matrix $\pmb{\Phi}$ \\
        $j_r$ & Number of components in the design matrix $\mathbf{R}$ \\
        $\mathbf{X}$ & Design matrix for PRF model, has shape ($l \times n$, $j_\phi \times j_r$)\\ 
        $\mathbf{B}$ & Matrix that used to create $\mathbf{f}_b$\\ 
        $b$ & Number of cadences to bin in time\\
        $\mathbf{f}_b$ & Flux array, having been binned in time, has shape ($l \times n$, $m / b$)\\
        $\pmb{\delta X}$ & Matrix consisting of the basis spline components for motion model. Each column vector is one spline component of the $\pmb{\delta x}$ vector. Has shape ($l \times n$, $j_x$)\\
        $\pmb{\delta Y}$ & Matrix consisting of the basis spline components for motion model. Each column vector is one spline component of the $\pmb{\delta y}$ vector. Has shape ($l \times n$, $j_y$)\\
        $j_x$ & Number of spline components to build in the $\delta x$ dimension for modeling motion\\
        $j_y$ & Number of spline components to build in the $\delta y$ dimension for modeling motion\\
        $\mathbf{A}_0$ & Matrix to build the time dependent model. Consists of the multiplication of every column in $\pmb{\delta X}$ with $\pmb{\delta Y}$. Has shape ($l \times n$, $j_x \times j_y$)\\    
        $\mathbf{A}$ & Matrix to build the time dependent model. Consists of $\mathbf{A}_0$, multiplied by . Has shape ($l \times n$ $\times$ $m$/$b$, $j_x \times j_y$ $\times$ 4)\\    
        $\mathbf{t}$ & Vector of times for each image in the image stack, has length $m$\\
        $\mathbf{t}_b$ & Vector of times for each image in the image stack having been mean binned, has length $b$\\
        $\mathbf{\hat{u}}$ & Vector of the best fit coefficients for the time dependent model ($\mathbf{A}$)\\
        $\mathbf{l}$ & Vector of the motion model in each pixel \\
    \end{tabular}
    \caption{All parameters described in this paper, and their meanings}
    \label{tab:nomenclature}
\end{table}

\section{Linear Least Squares}
\label{sec:linearappendix}

In this paper, we solve to find a best fitting model using linear least squares. In general, if we have a vector $\mathbf{y}$ containing data we assume that it is Gaussian distributed 
\begin{equation}
    \mathbf{y} \sim \mathcal{N}(\mathbf{m},\,\mathbf{K_y})\,
\end{equation}
where $\mathbf{K_y}$ is the covariance of the data and $\mathbf{m}$ is a vector with the same length as $\mathbf{y}$ and is the mean. We model $m$ as
\begin{equation}
    \label{eq:mod0}
    \mathbf{m} = \mathbf{X}\cdot\mathbf{w}
\end{equation}
where $\mathbf{X}$ is a design matrix, and $\mathbf{w}$ is a vector of weights. We can find the best fitting weights for a the design matrix by solving the linear system


\begin{equation}
    \label{eq:sigmainv0}
    \mathbf{K_w}^{-1}= \mathbf{X}^\intercal \cdot \mathbf{K}_y^{'-1} \cdot \mathbf{X} + \mathbf{C^{-1}}
\end{equation}
\begin{equation}
    \label{eq:what0}
    \mathbf{\hat{w}} = \mathbf{K_w}^{-1} \cdot \left(\mathbf{X}^\intercal \cdot \mathbf{K}_y^{'-1} \cdot \mathbf{y} + \mathbf{C^{-1}}.\pmb{\mu} \right).
\end{equation}
where 
\begin{equation}
     \mathbf{K}_y^{'-1} = (\mathbf{K_y} \circ \pmb{\Lambda}^{-1})^{-1}
\end{equation}
and
\begin{equation}
     \mathbf{C^{-1}} = \mathbf{I}.\left(\frac{1}{\pmb{\sigma}^2}\right)
\end{equation}

$\mathbf{\hat{w}}$ is a vector of the estimates of the best fitting coefficients. $\mathbf{K_y}$ is the covariance matrix of the data. $\pmb{\Lambda}$ is an optional diagonal matrix, consisting of weights for each element in $\mathbf{y}$. Setting $\pmb{\Lambda}$ to the identity matrix sets all pixels to equal weight, setting $\pmb{\Lambda}$ to a diagonal matrix of ones and zeros (or small values) allows some elements of $\mathbf{y}$ to be effectively excluded (masked) from the fit. $\pmb{\mu}$ is the prior mean of each of the $\mathbf{w}$, and $\pmb{\sigma}$ is the prior standard deviation of $\mathbf{w}$. $\mathbf{\hat{w}}$ has length equal to the the number of components in the design matrix. $\mathbf{K_w}$ is the covariance matrix of $\mathbf{w}$. We solve this system several times in this work. In practice, rather than using $\pmb{\Lambda}$ to change the weight of each element in the fit, we mask out elements of $\mathbf{y}$ and $\mathbf{K_y}$ in our Python implementation, using \texttt{numpy} arrays.

\section{Constructing B-Splines}
\label{sec:bsplines}

Basis splines, or B-splines, provide a way for us to build a simple linear model of piece-wise polynomials, which are forced to vary smoothly across all pieces. In this work, we use B-splines to model several aspects of the data. Our use of B-splines converts vector into a matrices, where each column of the matrix is a spline component. 

The B-spline polynomial pieces are defined by "knots". Each spline component is zero outside of a given pair of knots, and valued between knots. Splines are also defined by their degree, which we will denote as $k$. 

For example, in Section ~\ref{sec:linearmodel} we build a matrix of spline components from a vector $\mathbf{r}$ (the radial distance from a source). We specify knots $r_0, r_1, r_2, ..., r_{n_k}$, which are single values, and all lie between the lowest value of $\mathbf{r}$ and the highest value of $\mathbf{r}$. All the knots must been in ascending order. Depending on the degree, knots at the beginning and the end of the sequence must be repeat values. We specify a total of $n_k$ knots. There are $n_\mathbf{r}$ elements in the vector $\mathbf{r}$.

For a degree 1 b-spline, a single spline component (column vector) for the $i$th knot is given as
\begin{equation}
\mathbf{b}_{i, 1} = \begin{cases}
        1 & \; \textrm{if} \;r_{i, 1} \leq \mathbf{r} < r_{i + 1, 1}\\
        0 & \textrm{otherwise}
        \end{cases}
\end{equation}

The matrix defining the b-spline model is then
\begin{equation}
    \mathbf{R}_1 = \begin{bmatrix}
    \mathbf{b}_{0, 1} & \mathbf{b}_{1, 1} & \mathbf{b}_{2, 1} & ... & \mathbf{b}_{n, 1} 
    \end{bmatrix}
\end{equation}
where each $\mathbf{b}_i$ is a column vector of length $n_\mathbf{r}$, and so $\mathbf{R}_1$ is a matrix, with shape ($n_\mathbf{r}$, $n_k$). Each $\mathbf{b}_i$ is a step function, which is 1 between knots. 

B-splines of order $k>$1 are defined by a recursive relation. Each vector of the second order B-spline is given as

\begin{equation}
    \mathbf{b}_{i, 2} = \pmb{\omega}_{i, 1} \circ \mathbf{b}_{i, 1} + 
    (1 - \pmb{\omega}_{i + 1, 1}) \circ  \mathbf{b}_{i + 1, 1}
\end{equation}
where $\circ$ indicates the element-wise product and
\begin{equation}
    \pmb{\omega}_{i, k} = \begin{cases}
    \frac{\mathbf{r} - r_i}{r_{i + k} - r_i} & r_{i + k} \neq r_i\\
    0 & \textrm{otherwise}
    \end{cases}
\end{equation}

This creates a set of piecewise, 1st order polynomials between each set of knots. The second order basis spline matrix $\mathbf{R}_2$ is given by the matrix of all column vectors for every $i$.

Higher orders of the b-spline can be made recursively using 

\begin{equation}
    \mathbf{b}_{i, k + 1} = \pmb{\omega}_{i, k} \circ \mathbf{b}_{i, k} + 
    (1 - \pmb{\omega}_{i + k, 1}) \circ  \mathbf{b}_{i + 1, k}.
\end{equation}

\begin{figure*}
    \centering
    \includegraphics[width=\textwidth]{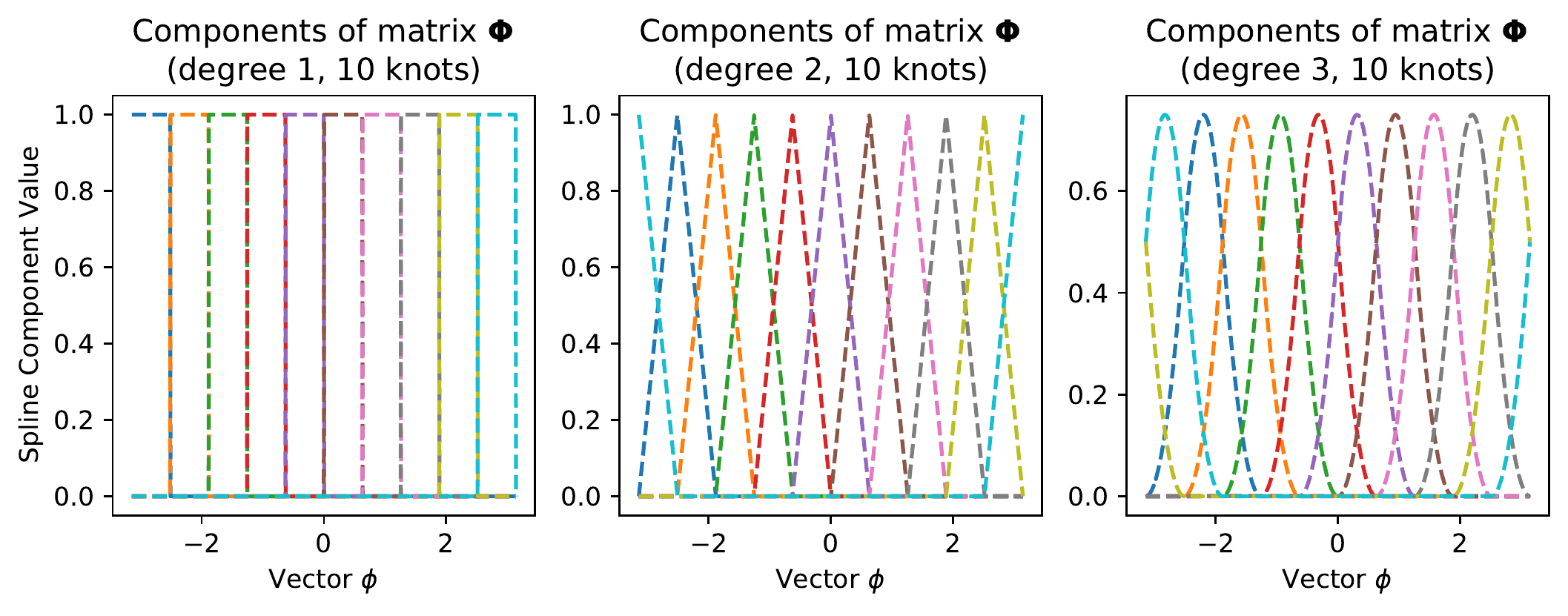}
    \caption{Example of a the spline matrix $\pmb{\Phi}$ used in this work, at different degrees. In this work we use a 3rd degree spline. This spline has been created using the "wrapped" spline code given in Appendix~\ref{sec:wrap}, and so the value at $\pi$ is the same as the value at $\pi$. B-splines are broken into piecewise polynomials. Increasing the degree of the spline increases the order of the polynomial in each piece. Fitting $\pmb{\Phi}$ to a vector using the equations in Appendix~\ref{sec:linearappendix} will result in a smooth trend.}
    \label{fig:spline}
\end{figure*}

In this work, we find a 3rd order B-spline adequately trades model complexity with efficient computing, and employ $k=3$ throughout this work. The above proceedure is used in this work to create B-spline matrices $\pmb{\delta X}$, $\pmb{\delta Y}$, $\pmb{\Phi}$, $\mathbf{R}$ from vectors $\pmb{\delta x}$, $\pmb{\delta y}$, $\pmb{\phi}$, $\mathbf{r}$. We direct the reader to Appendix~\ref{sec:wrap} for an example of a Python implementation of a B-spline that wraps at a value of $\pi$, such that the value at $-\pi$ is equal to the value at $\pi$, which we use to create the matrix $\pmb{\Phi}$. An example of the matrix structure obtained from an input vector is shown in Figure~\ref{fig:spline}.

\section{Wrapped Spline Python Implementation}
\label{sec:wrap}
\begin{minted}{python}
import numpy as np
def wrapped_spline_matrix(input_vector, degree=2, nknots=10):
    """
    Creates a matrix of splines according to the input vector. This will wrap between -pi and pi.
    This is meant to be used to build the basis vectors for periodic data, like the angle in polar
    coordinates.

    Parameters
    ----------
    input_vector : numpy.ndarray
        Input data to create basis, angle values MUST BE BETWEEN -PI and PI.
    degree : int
        Degree of the spline basis
    nknots : int
         Number of knots for the splines
    Returns
    -------
    folded_basis : numpy.ndarray
        Array of folded-spline basis
    """
    if (degree <= 0) | (not isinstance(degree, int)):
      raise ValueError('Degree must be an integer greater than 0')

    if not ((input_vector >= -np.pi) & (input_vector <= np.pi)).all():
        raise ValueError("Must be between -pi and pi")
    x = np.copy(input_vector)
    # in order to wrap, we'll evaluate between -pi and 3pi 
    x1 = np.hstack([x, x + np.pi * 2])
    nt = (nknots * 2) + 1

    t = np.linspace(-np.pi, 3 * np.pi, nt)
    dt = np.max(np.diff(t))

    # Degree 1 basis
    basis = np.asarray(
        [
            ((x1 >= t[idx]) & (x1 < t[idx + 1])).astype(float)
            for idx in range(len(t) - 1)
        ]
    )
 
    # Higher order basis
    for order in np.arange(1, degree):
        basis_1 = []
        for idx in range(len(t) - 1):
            a = ((x1 - t[idx]) / (dt * order)) * basis[idx]

            if ((idx + order + 1)) < (nt - 1):
                b = (-(x1 - t[(idx + order + 1)]) / (dt * order)) * basis[
                    (idx + 1) % (nt - 1)
                ]
            else:
                b = np.zeros(len(x1))
            basis_1.append(a + b)
        basis = np.vstack(basis_1)

    folded_basis = np.copy(basis)[: nt // 2, : len(x)]
    for idx in np.arange(-(degree), 0):
        folded_basis[idx, :] += np.copy(basis)[nt // 2 + idx, len(x) :]
    return folded_basis
    
\end{minted}

\facilities{Kepler}
\software{lightkurve \citep{lightkurve}, exoplanet \citep{exoplanet:exoplanet}, astropy \citep{astropy:2013, astropy:2018}, scipy \cite{scipy}, PSFMachine \citep{PSFMachine}}

\bibliographystyle{aasjournal}
\bibliography{bib.bib}

\end{document}